\documentclass[
superscriptaddress,
 amsmath,amssymb,
 aps,
 lengthcheck,%
]{revtex4}

\usepackage{graphicx}
\usepackage{dcolumn}
\usepackage{bm}
\usepackage{hyperref}
\usepackage{feynmp}
\usepackage{amsmath}

\begin{document}

\title{Excitation Spectra of Bosons in Optical Lattices from Schwinger-Keldysh
Calculation}

\author{T. D. Gra\ss}
\affiliation{ICFO-Institut de Ci{\`e}ncies Fot{\`o}niques, Mediterranean Technology Park, 08860 Castelldefels (Barcelona),
Spain}
\author{F. E. A. dos Santos}
\affiliation{Institut f\"ur Theoretische Physik, Freie Universit\"at Berlin, Arnimallee 14, 14195 Berlin, Germany}
\author{A. Pelster}
\affiliation{Fachbereich Physik, Universit\"at Duisburg-Essen, Lotharstra\ss e 1, 47048 Duisburg, Germany}

\date{\today}

\begin{abstract}
Within the Schwinger-Keldysh formalism we derive a Ginzburg-Landau theory for the Bose-Hubbard model which describes 
the real-time dynamics of the complex order parameter field. Analyzing the excitations in the vicinity of the quantum 
phase transition it turns out that particle/hole dispersions in the Mott phase map continuously onto corresponding 
amplitude/phase excitations in the superfluid phase, which have
been detected recently by Bragg spectroscopy measurements.
\end{abstract}


\maketitle

\newcommand{\op}[1]{\ensuremath{\Hat{\mathrm{#1}}}}
\newcommand{\setval}{\fmfset{wiggly_len}{2.5mm}\fmfset{arrow_len}{2.5mm}\fmfset{arrow_ang}{13}\fmfset{dash_len}{2.5mm}
\fmfpen{0.18mm}\fmfset{dot_size}{1thick}}
\setlength{\unitlength}{1mm}

\section{Introduction \label{intro}}

Within the last decade ultracold atoms in optical lattices \cite{jaksch,bloch1} have become a standard tool for studying 
quantum-statistical many-body effects. Due to their high tunability, these systems represent an almost perfect test 
ground for a large variety of solid-state models. In particular, the experimental observation of the seminal quantum 
phase transition from the Mott insulating (MI) to the superfluid (SF) phase, exhibited by a single-band Bose-Hubbard 
(BH) system of spinless or spin-polarized bosons, has recently attracted a lot of attention 
\cite{bloch-review,latticereview}. 
Although the occurrence of this quantum phase transition is evident from the momentum distributions of 
time-of-flight measurements, its precise location cannot be determined from them. Recently, 
however, more detailed information about the collective excitations of this system could also be achieved by exciting 
the system via lattice modulation \cite{esslinger} or by Bragg spectroscopy
\cite{sengstock,clement,gapped-mode}. Deep in the SF phase, the 
observed gapless excitation spectrum can be well understood within a Bogoliubov theory \cite{stoof}. Approaching the 
phase boundary, a time-dependent dynamic Gutzwiller calculation \cite{gapped-mode}, 
a slave-boson method \cite{blatter1}, and a random-phase
approximation \cite{menotti} have predicted an additional SF gapped mode, which
recently could be confirmed experimentally in the strongly
interacting regime \cite{gapped-mode}. An open question is the fate of this
mode in the weakly interacting Bogoliubov limit. When the MI phase is reached,
both SF modes turn continuously into particle and 
hole excitations which are also found by mean-field theory \cite{stoof}. Due to
their finite energy gaps, they characterize the 
insulating phase in a unique way.

A field-theoretic ansatz describing the system in both the insulating and the
superfluid regime has first been considered 
in Ref.~\cite{dupuis}, where an effective action has been obtained via two successive Hubbard-Stratonovich 
transformations. The same action can also be obtained by the Ginzburg-Landau approach developed in 
Refs.~\cite{ednilson,barry}, which is technically based on resumming a perturbative hopping expansion. As is explicitly 
shown in Refs.~\cite{ednilson,barry,eckardt,holthaus1,holthaus2}, 
the practical advantage of this non-perturbative approach is that it provides a generalization
of mean-field theory by taking into account higher hopping orders in a straight-forward systematic way. 
However, the Landau theory Ref.~\cite{ednilson} is restricted to a static 
description of the SF-MI transition at zero temperature. A finite-temperature Ginzburg-Landau
theory developed in Ref.~\cite{barry} 
yields also dynamic results via an analytic continuation, but is restricted to near-equilibrium situations. Since 
time-resolved measurements \cite{Collapse1} have become possible within the last years, a real-time description of 
quantum systems is desirable. Therefore, we follow Ref. \cite{laserphysics} and 
modify the imaginary-time approach from Ref.~\cite{barry} by converting
it to real time,  thus requiring techniques which were 
first introduced by L.V. Keldysh \cite{keldysh} and J. Schwinger \cite{schwinger}.

To this end we start
with a brief introduction to the so-called Schwinger-Keldysh formalism of real-time 
perturbation theory at finite temperature in Section \ref{keldysh}. The actual Ginzburg-Landau theory 
for the underlying Bose-Hubbard model is then derived  
by adapting the non-perturbative imaginary-time calculation of Ref.~\cite{barry} to the Keldysh space.
Thus, we combine a perturbative hopping expansion of the free energy
in Section \ref{per} with a subsequent resummation due to a Legendre transformation in Section \ref{resum}
and determine the resulting effective action. 
Its extremization leads to equations of motion for the order parameter fields
which are solved
in Section \ref{sem}, yielding a plethora of physical results like the phase boundary 
and the excitation spectra in Section \ref{res}. Furthermore, we compare in Section \ref{com}
the present Schwinger-Keldysh calculation with the corresponding imaginary-time approach of Ref.~\cite{barry}, 
revealing an unexpected mismatch between both descriptions. Finally, a summary
of the present work,
which is restricted to equilibrium physics, and 
an outlook, how to generalize it to non-equilibrium situations, is given in Section \ref{sum}. 
\section{Schwinger-Keldysh Formalism \label{keldysh}}
In order to introduce the Schwinger-Keldysh formalism, we consider the Green
functions of a many-body system. 
In general, these are averages over a product of creation operators $\op{a}^\dagger_i$ and annihilation operators 
$\op{a}_i$, where the index $i$ collects the degrees of freedom of the system. In our case, it just 
denotes the respective lattice site. 
At zero temperature, the averaging process is purely quantum mechanical, whereas for finite temperature also the thermal 
fluctuations are taken into account. In equilibrium systems, i.e. those with a Hamiltonian constant in time, introducing 
imaginary-time variables puts the thermal averaging formally on an equal footing 
with the dynamic evolution of a time-dependent system at zero temperature.

For both cases, perturbative treatments are well known, which are based on reducing the non-trivial
real or imaginary dynamics of the 
operators in the Heisenberg picture to the trivial dynamics of operators in the Dirac picture, which is determined by some 
solvable part $\op{H}_0$ of the full Hamiltonian $\op{H}$. The remaining part $\op{H}_1$ 
then enters the time-evolution operator $\op{U}(t,t_0)$ which acts 
on the states. In the case of a real-time evolution, for 
instance, it reads $\op{U}(t,t_0) = \op{T} \exp \left[ -\frac{i}{\hbar} \int_{t_0}^t \mathrm{d}t_1 \op{H}_{1\mathrm{D}}
(t_1)\right]$, where $\op{T}$ is the time-ordering operator. 

In the zero-temperature formalism (ZTF), the Green functions are defined as
ground-state expectation values of 
time-ordered operator products in the Heisenberg picture, i.e. $\Big\langle
\op{T} \left[ \op{a}_{i_1\mathrm{H}}^\dagger
(t_1) \cdots \op{a}_{i_n\mathrm{H}}(t_n) \right] \Big\rangle_{\op{H}}$, where
$\langle \bullet \rangle_{\op{H}}$ denotes the 
expectation value with respect to the ground-state of the Hamiltonian $\op{H}$.
Applying the translation formula from 
the Heisenberg to the 
Dirac picture to this expression, i.e. $\op{a}_{\rm{H}}(t) = \op{U}(t_0,t)
\op{a}_{\rm{D}}(t) \op{U}(t,t_0)$, 
where $t_0$ is the initial time, we find for each operator a forward and a
backward time-evolution. But due to the 
time-ordering, these various pieces can be straightened to one single forward
evolution from $t_0$ to the largest time 
within the operator product and a subsequent backward evolution. The latter,
however, cancels due to 
a connection between the 
ground-states of the full and the unperturbed system, 
which is provided by the Gell-Mann-Low theorem \cite{gell-mann}. In this way, 
we do not only get rid of the backward evolution, but also reduce the
expectation value to the one of the known, 
unperturbed ground state. Finally we get $\Big\langle \op{T} \left[
\op{U}(\infty,-\infty) 
\op{a}_{i_1\mathrm{D}}^\dagger(t_1) \cdots \op{a}_{i_n\mathrm{D}}(t_n) \right]
\Big\rangle_{\op{H}_0}$, where the 
non-trivial part of the averaging process enters  only via a forward evolution
along a straight time path, which has 
been extended to infinity. In this way, the perturbation theory is reduced to a
Taylor expansion of the time-evolution 
operator $\op{U}(\infty,-\infty)$.

In the following, however, we also consider finite temperatures, so in the
Green functions the average 
$\langle \bullet \rangle_{\op{H}}$ now denotes a thermal average with respect to
the equilibrium of $\op{H}$, i.e.
$\mathrm{Tr}\left(\bullet \,\,\mathrm{e}^{-\beta \op{H}} 
\right)/\mathrm{Tr}\left(\mathrm{e}^{-\beta \op{H}}\right)$, where 
$\beta= 1/k_{\rm B}T$ is as usual 
the inverse temperature. If there is no evolution along the real-time axis, then
the scheme of the ZTF 
described above can be taken over to an artificially introduced imaginary-time
axis. Evolution along this axis somehow 
imitates the thermal averaging process, so the backward evolution in imaginary
dynamics of the operator product can 
directly be cancelled against the perturbative part of the thermal density
matrix $\mathrm{e}^{-\beta \op{H}}$. In this 
way, the imaginary-time formalism (ITF) also puts all non-trivial $\op{H}_1$
contributions of the Green functions 
into a straight time evolution.

\begin{figure}
\center
\includegraphics[width=8cm]{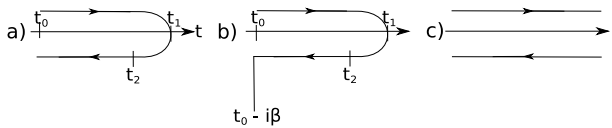}
\caption{Contours of time evolution: a) Schwinger contour, b) interaction
contour, c) Keldysh contour.}
\label{cont}
\end{figure}

The situation becomes more complicated for a time-dependent many-body system at
finite temperature where the backward evolution along the real-time axis
cannot be canceled by the Gell-Mann-Low theorem as in the ZTF. To realize the 
consequences of this, let us consider for simplicity the product
$\op{O}_{2\mathrm{H}}(t_2)\op{O}_{1\mathrm{H}}(t_1)$ of two 
arbitrary operators in the Heisenberg picture. If $t_1>t_2$, the translation
from the Heisenberg into the Dirac picture 
yields
$\op{U}(t_0,t_2)\op{O}_{2\mathrm{D}}(t_2)\op{U}(t_2,t_1)\op{O}_{1\mathrm{D}}
(t_1)\op{U}(t_1,t_0)$, which represents
a time-evolution along a closed contour as depicted in Fig.~\ref{cont} a) with
$\op{O}_{2\mathrm{D}}(t_2)$ being located on the 
backward path of the contour. For the opposite case $t_2 > t_1$, the closed
contour extends from $t_0$ to $t_2$, and the condition to maintain the
order of the operators now is that $\op{O}_{1\mathrm{D}}(t_1)$ appears on the
forward path. In both cases it plays obviously no role, which part of the 
contour is assigned to the last operator.

By providing each operator with a corresponding 
path index $\mathrm{P}=\pm$, where $+$ stands for the forward and $-$ for the
backward 
path, the positions of the operators on the contour can be fixed. With this 
we are able to define a contour-ordering operator 
$\op{T}_\mathrm{c}$, which brings  the $+$operators in time order, while acting
as an anti-time ordering operator 
$\op{A}$ on the $-$operators. Such contour-ordered operator products are the
most natural generalization of the 
time-ordered products considered in the ZTF. Thus the relevant Green functions
are given by quantities like
$\Big\langle \op{T}_\mathrm{c} \left[\op{a}_{i_1\mathrm{H}}^{\mathrm{P}_1
\dagger}(t_1) \cdots 
\op{a}_{i_n\mathrm{H}}^{\mathrm{P}_n}(t_n) \right] \Big\rangle_{\op{H}}$. But
even after transforming into the Dirac picture,
i.e. $\Big\langle \op{T}_\mathrm{c} \left[ \op{U}(t_0,t_>)\op{U}(t_>,t_0) \
\op{a}_{i_1\mathrm{D}}^{\mathrm{P}_1 \dagger}(t_1) 
\cdots \op{a}_{i_n\mathrm{D}}^{\mathrm{P}_n}(t_n) \right] \Big\rangle_{\op{H}_0}$,
where $t_>$ is the biggest time appearing in 
the operator product, the perturbative part of the Hamiltonian is not completely
isolated within the time-evolution 
operator, since it still appears in the thermal average. We might manage this by
extending the time-evolution contour to 
the complex plane as shown in Fig.~\ref{cont} b), yielding a third path.

However, it is widely believed in the literature that, if we perform 
the initial time limit $t_0\rightarrow-\infty$, the imaginary 
part of that contour can be neglected for initially uncorrelated systems
\cite{chou,rammer-buch}. We will follow this 
tradition and discuss its limitations at the end of this paper. With this
simplification and by extending the contour to 
the infinite future, we get the time-evolution contour from Fig.~\ref{cont} c),
encircling completely the real-time axis. 
It is a closed path, and we, therefore, refer to this formalism as the closed
time path formalism (CTPF).

Then the contour-ordered Green functions can be defined as
\begin{eqnarray}
\label{HD} 
&& G_{\{i\}}^{\{\mathrm{P}\}}(\{t\}) \equiv i^{n+m-1} \Big\langle
\op{T}_\mathrm{c} \left[ \op{S}^\dagger \op{S}  \nonumber 
\ \op{a}_{i_1}^{\mathrm{P}_1}(t_1)\cdots \op{a}_{i_n}^{\mathrm{P}_n}(t_n)
\right.
\\ && \left.
\times \op{a}_{i_{n+1}}^{\mathrm{P}_{n+1} \dagger}(t_{n+1}) \cdots
\op{a}_{i_{n+m}}^{\mathrm{P}_{n+m} \dagger}(t_{n+m}) \right] 
\Big\rangle_{\op{H}_0},
\end{eqnarray}
where we have waived for simplicity
the index for the Dirac picture, as we will do so for the rest of this paper.
With the curly brackets 
$\{ X \}$, we abbreviated the union of variables $X_1,\cdots,X_{n+m}$. In this
expression, the perturbation enters only 
in the time evolution $\op{S}^\dagger\op{S}=\op{T}_{\mathrm{c}} \exp \left\{
-\frac{i}{\hbar} \int_{-\infty}^{\infty} 
\mathrm{d}t \left[  \op{H}_{1}^+(t) - \op{H}_{1}^-(t) \right] \right\}$.

The Green functions from Eq.~(\ref{HD}) can systematically be derived from a
generating functional. To this end, we have 
to introduce artificial currents $j^{\mathrm{P}}_i(t)$ and their complex conjugate
$\bar{j}^{\mathrm{P}}_i(t)$, which linearly couple to the 
corresponding creation and annihilation operators
$\op{a}_i^{\mathrm{P}\dagger}(t)$ and $\op{a}^{\mathrm{P}}_i(t)$. With 
such a source term, the new Hamiltonian reads $\op{H}^{\mathrm{P}}[j,\bar j](t)
= \op{H}_0^{\mathrm{P}}(t) 
+ \op{H}^{\mathrm{P}}_1(t) + \sum_i \left[ j_i^{\mathrm{P}}(t)
\op{a}_i^{\mathrm{P}\dagger}(t) + \bar j_i^{\mathrm{P}}(t) 
\op{a}_i^{\mathrm{P}}(t) \right]$.
Assigning the source term to the perturbative part, we have to substitute
$\op{S}$ by $\op{S}[j^+,\bar j^+] = \op{T} \exp 
\Big( -\frac{i}{\hbar} \int_{-\infty}^{\infty} \mathrm{d}t \big\{
\op{H}_{1}^+(t) + \sum_i \big[ j_i^+(t) 
\op{a}_i^{+\dagger}(t) +\mathrm{c.c.} \big] \big\} \Big)$, and similarly for the
backward evolution operator 
$\op{S}^\dagger$. The generating functional is then defined as
\begin{eqnarray}
\label{Z}
 {\cal Z}[j^+,j^-,\bar j^+, \bar j^-] = \left\langle \op{T}_{\mathrm{c}} \{
\op{S}^\dagger[j^-, \bar
j^-] \op{S}[j^+, \bar j^+] \} 
\right\rangle_{\op{H}_0},
\end{eqnarray}
yielding the Green functions as its functional derivatives
\begin{eqnarray}
\label{GD}
 G_{i_1\cdots i_n,i_{n+1}\cdots i_{n+m}} = \frac{\delta^{n+m} {\cal Z}[j,\bar
j]}{\delta \bar j_{i_1} \cdots \delta \bar j_{i_n} 
\delta j_{i_{n+1}} \cdots \delta j_{i_{n+m}}} \Bigg|_{j=0}^{\bar j =0},
\end{eqnarray}
where both path indices and temporal variables have been waived for simplicity.

Path-ordered quantities, though naturally arising within the Keldysh formalism,
lack a clear physical interpretation, so it is convenient to define the linear
combinations 
\begin{eqnarray}
\label{k-base1}
X^{\Sigma}(t) &=& [X^+(t) + X^-(t)]/\sqrt{2} \, , \\
\label{k-base2}
X^{\Delta}(t) &=& [X^+(t) - X^-(t)]/\sqrt{2}\,.
\end{eqnarray} 
A good discussion of different choices for a suitable basis
can be found in Ref.~\cite{weert}.
To see how the path-ordered Green functions
given in Eq.~(\ref{HD}) transform under this so-called Keldysh rotation, it is
most instructive to consider the 2-point functions. Writing them in form of a
2x2 matrix, the transformation is given by
\begin{align}
\label{krot}
 \left(
 \begin{array}{cc}
  G^{++}_{ij}(t;t') & G^{+-}_{ij}(t,t') \\
  G^{-+}_{ij}(t;t') & G^{--}_{ij}(t,t') \\
 \end{array}
 \right)
\rightarrow
\left( \begin{array}{cc}
A_{ij}(t;t') & G_{ij}^{\mathrm{R}}(t;t')  \\
G_{ij}^{\mathrm{A}}(t;t') & 0  \\
\end{array} \right),
\end{align}
where $G^\mathrm{R,A}$ are the well known retarded/advanced Green function and
$A$ is the anticommutator function. 
In the case of bosons, they are defined as \cite{kad}
\begin{eqnarray}
\label{gret}
 G^{\mathrm{R}}_{ij}(t;t') &\equiv& i\theta(t-t') \Big\langle
\left[\op{a}_{i\mathrm{H}}(t), \op{a}^\dagger_{j\mathrm{H}}(t') 
\right]\Big\rangle_{\op{H}}, \\
\label{gav}
 G^{\mathrm{A}}_{ij}(t;t') &\equiv& -i\theta(t-t') \Big\langle
\left[\op{a}_{i\mathrm{H}}(t), \op{a}^\dagger_{j\mathrm{H}}(t')
 \right]\Big\rangle_{\op{H}}, \\
\label{ga}
 A_{ij}(t;t') &\equiv& i\Big\langle \left[\op{a}_{i\mathrm{H}}(t),
\op{a}^\dagger_{j\mathrm{H}}(t') \right]_+
\Big\rangle_{\op{H}}.
\end{eqnarray}
Here, $[\bullet,\bullet]_+$ 
denotes the anticommutator, while the commutator is given by
$[\bullet,\bullet]$, and $\theta$  
stands for the Heaviside function. To see how these quantities arise from the
Keldysh rotation in Eq.~(\ref{krot}), we must 
note that the definitions in Eqs.~(\ref{gret})--(\ref{ga}) can be rewritten in
terms of contour-ordered products 
$\op{T}_\mathrm{c} \big[ \op{a}_{i\mathrm{H}}^{\mathrm{P}}(t)
\op{a}^{\mathrm{P}'\dagger}_{j\mathrm{H}}(t') \big]$. 
It then turns out that 
these can be summarized to one single product of $\op{a}^\Sigma$ and
$\op{a}^\Delta$ operators or their hermitian 
conjugates, respectively. We find:
\begin{eqnarray}
 G^{\mathrm{R}}_{ij}(t;t') = i \Big\langle \op{T}_\mathrm{c}\big[
\op{a}_{i\mathrm{H}}^\Sigma(t) \op{a}_{j\mathrm{H}}^{\Delta 
\dagger}(t') \big] \Big\rangle_{\op{H}}, \\
 G^{\mathrm{A}}_{ij}(t;t') = i \Big\langle \op{T}_\mathrm{c}\big[
\op{a}_{i\mathrm{H}}^\Delta(t) \op{a}_{j\mathrm{H}}^{\Sigma 
\dagger}(t') \big] \Big\rangle_{\op{H}}, \\
 A_{ij}(t;t') = i \Big\langle \op{T}_\mathrm{c}\big[
\op{a}_{i\mathrm{H}}^\Sigma(t) \op{a}_{j\mathrm{H}}^{\Sigma \dagger}(t') 
\big] \Big\rangle_{\op{H}}.
\end{eqnarray}
Since in the next section we are going to calculate the path-ordered Green
functions perturbatively, we stress that 
these relations do not only hold for the exact Green functions, but also in
any order of perturbation theory 
\cite{kamenev2}.
Furthermore, it can be shown  for arbitrary $n$-point functions that the Keldysh
rotation always yields, 
amongst other functions, the retarded and advanced Green function. For
practical calculations it is also relevant 
to notice that operator products of the kind 
$\op{T}_{\mathrm{c}} \op{O}_1^\Delta \cdots
\op{O}_n^\Delta$ are identically zero \cite{chou}, 
so amongst the Keldysh-rotated $n$-point functions, we always find a vanishing
one.
\section{Perturbation Theory \label{per}}
In the previous section we have introduced the general formalism of separating
the non-trivial real-time dynamics and 
thermodynamics from the rest of a Green function by performing a translation
from the Heisenberg to the 
Dirac picture. In view of concrete applications, we must now define the
respective
unperturbed and perturbed parts of the considered Hamiltonian.

Bosons in an optical lattice are described by
the BH-Hamiltonian  \cite{jaksch,bloch1} given as $\op{H}_{\mathrm{BH}} =
\op{H}_0 + \op{H}_1$, where we have introduced a decomposition into
the local part
\begin{eqnarray}
\label{H0}
\op{H}_0 = \sum_{i} \left[ \frac{U}{2} \op{a}_i^{\dagger} \op{a}_i \left(
\op{a}_i^{\dagger} \op{a}_i -1 \right) - \mu 
\op{a}_i^{\dagger} \op{a}_i \right]
\end{eqnarray}
and the non-local hopping term
\begin{eqnarray}
\op{H}_1= - \sum_{i,j} J_{ij} \op{a}_i^{\dagger} \op{a}_j.
\end{eqnarray}
Here $\op{a}_i$ ($\op{a}^\dagger_i$) denotes the bosonic annihilation (creation)
operator at lattice site 
$i$, $\mu$ the chemical potential, $U$ the on-site interaction parameter, and
$J_{ij}$ the hopping matrix element 
being equal to $J>0$ for nearest neighbors only.

The main feature of this Hamiltonian is a quantum phase transition from a MI
phase to a SF phase due to the 
competition between kinetic energy, i.e. hopping between sites, and local
on-site interactions. Therefore, it is natural 
to consider the hopping as a perturbation, if we are interested in the MI
phase, whereas in the SF phase the
interaction represents the perturbation.
However, a perturbative description of the critical behavior in between both
regimes would then be doomed 
to fail. But we are in the lucky situation that a \textit{resummed} hopping
expansion can also be considered as a 
$1/D$-expansion of a $D$-dimensional system and, thus, might allow for a proper
description of the quantum phase transition. 
To this end we note that, in 
the presence of a condensate, the hopping parameter $J$ must be rescaled according to the dimensional scaling
law $J\rightarrow J/D$ in order to have a finite energy in infinite
dimensions \cite{vollhardt}. This means that 
within a hopping expansion all $n$th order hopping loops are suppressed by a
factor $1/D^{n-1}$. By resumming the 
1-particle irreducible contributions up to the $n$th hopping order, we therefore
get an effective $(1/D)$-expansion up 
to the $(n-1)$th order. Although we restrict ourselves 
in this paper to the lowest order $n=1$, we will get a theory
which 
is exact for infinite dimensions or infinite-range hopping \cite{fisher}. 
In this way we find for $D \geq 2$ that the quantum 
phase transition can be well described in a quantitative way.

Finally we must add a source term to the Hamiltonian, yielding $\op{H}[j, \bar
j](t) = \op{H}_{\mathrm{BH}}
+ \op{H}_{\mathrm{S}}(t)$, with
\begin{eqnarray}
\label{source}
\op{H}_{\mathrm{S}}[j,\bar j](t)= \sum_i \left[ \op{a}_i^\dagger(t) j_i(t) +
\mathrm{c.c.} \right] \,.
\end{eqnarray}
A similar source term has already been introduced for defining a generating functional in
Eq.~(\ref{Z}), but there it has 
only been a technical tool. Although we will set the currents to zero in the
end, here this term acquires a 
physical meaning, since it explicitly
breaks the underlying $U(1)$ symmetry of the original BH-Hamiltonian. Only this
symmetry-breaking makes 
it possible to have a non-vanishing condensate amplitude $\langle \op{a}_i(t)
\rangle$, which characterizes the SF phase 
and can be taken as an order parameter field $\Psi_i(t)$ for describing the
MI-SF transition within a Ginzburg-Landau 
theory. From 
the definition (\ref{source}) it is clear that the currents $j_i(t)$ and
$\bar{j}_i(t)$
are the conjugate variables of the order parameter fields 
$\bar{\Psi}_i(t) = \langle \op{a}^{\dagger}_i(t) \rangle$ and 
$\Psi_i(t) = \langle \op{a}_i(t) \rangle$, now appearing explicitly
within the Hamiltonian
$\op{H}[j,\bar j](t)$.

Ginzburg-Landau theories are in general based on a fourth-order expansion of the
thermodynamic potential 
in the order parameter field \cite{sachdev}. In order to derive it for the underlying Bose-Hubbard model,
we will start from  
the free energy. For small order parameter fields we can assume that also the
influence of the symmetry-breaking currents is small,
so we can write down the free energy as a power series
with respect to these 
currents $j_i(t)$ and $\bar{j}_i(t)$. It is then possible to
replace the currents via a 
Legendre transformation by the order parameter fields 
$\bar{\Psi}_i(t) = \langle \op{a}^{\dagger}_i(t) \rangle$ and 
$\Psi_i(t) = \langle \op{a}_i(t) \rangle$, so that 
we finally end up with the so-called effective action that serves as a
Ginzburg-Landau functional.

Noting that the generating functional ${\cal Z}[j^+,j^-,\bar j^+, \bar j^-]$
defined in
Eq.~(\ref{Z}) has a similar structure as the 
partition function in statistical mechanics, we get an analogue of the free
energy by taking the logarithm: 
\begin{eqnarray}
\label{F}
 {\cal F}[j^+, j^-, \bar j^+, \bar j^-] = -i \ln {\cal Z}[j^+, j^-, \bar j^+,
\bar j^-].
\end{eqnarray}
As described in Section \ref{keldysh}, the functional ${\cal Z}$ is
given by a forward and a backward evolution in time, where the operators and
currents appearing on both time paths have to be distinguished from each other.
Before starting a perturbative expansion of this functional, however, it turns
out to be feasible to perform a Keldysh rotation, which mixes the quantities on the
forward and backward path as defined by Eqs. (\ref{k-base1}) and
(\ref{k-base2}). Under this rotation, the source term transforms as $j_i^+
\op{a}_i^{+\dagger} - j_i^- \op{a}_i^{-\dagger} =
j_i^\Sigma \op{a}_i^{\Delta\dagger} + j_i^\Delta \op{a}_i^{\Sigma\dagger}$, and
for the hopping term we have
$\op{a}_i^{+\dagger} \op{a}_j^+ - \op{a}_i^{-\dagger}\op{a}_j^- =
\op{a}_i^{\Delta\dagger} \op{a}_j^{\Sigma} + \op{a}_i^{\Sigma
\dagger}\op{a}_j^{\Delta}$.
It turns out to be helpful to introduce the vector quantities
${\bf j}_i= (j_i^\Delta, j_i^\Sigma)^T$ and $\op{{\bf a}}_i^\dagger =
(\op{a}_i^{\Sigma\dagger},\op{a}_i^{\Delta\dagger})$. Then the generating
functional reads
\begin{widetext}
\begin{align}
\label{F1}
 {\cal F}[{\bf j},{\bf \bar j}] =  -i \ln \Bigg\langle \op{T}_{\mathrm{c}} \exp
\Big(
-\frac{i}{\hbar} \int_{-\infty}^{\infty} \mathrm{d}t \Big\{ \sum_{ij} J_{ij}
\op{{\bf a}}_j^\dagger \sigma^1 \op{{\bf a}}_i +
\sum_i \big[ \, {\bf \bar j}_i(t) \cdot
\op{{\bf a}}_i(t) +\mathrm{c.c.} \big] \Big\} \Big)
\Bigg\rangle_{\op{H}_0},
\end{align}
where $\sigma^1$ is the Pauli matrix \cite{rammer}.
\end{widetext}
Instead of straightforwardly expanding this functional in the hopping
parameter $J$ and in the vector currents ${\bf j}$ and ${\bf \bar j}$, we first
note that the logarithm has the nice property to make ${\cal
F}[{\bf j}, {\bf \bar j}]$ an extensive quantity. Thus the linked-cluster
theorem
\cite{singh,metzner} applies: It states that, whereas the expansion of
${\cal Z}$ is built up by the Green functions, which have been defined in the
path-ordered basis in Eq.~(\ref{HD}), the functional ${\cal F}$ can be
expanded in a series of \textit{connected} Green functions or cumulants $C$:
\begin{eqnarray}
\label{CD}
\hspace*{-2mm}  C_{i_1\cdots i_n;i_{n+1}\cdots i_{n+m}} = \frac{\delta^{n+m} {\cal F}[j,\bar
j]}{\delta \bar j_{i_n} \cdots \delta \bar j_{i_n} 
\delta j_{i_{n+1}} \cdots \delta j_{i_{n+m}}}\Bigg|_{j=0}^{\bar j = 0}.
\end{eqnarray}
For simplicity, we have suppressed here again both the time variables and the
Keldysh indices.

These cumulants are much simpler objects than the Green functions: Whereas
roughly speaking the Green functions may 
describe any particle creation and annihilation processes, the cumulants
decompose
them into their independent 
contributions. 
What is meant by this, can be illustrated by considering only the unperturbed
system $\op{H}_0$ in (\ref{H0}). 
Here, each lattice site $i$ 
can be considered as an independent system and we could define generating
functionals 
${\cal Z}^{(0)}_i[{\bf j}_i, {\bf \bar j}_i]$ in each 
subsystem. We would find that the Green functions of the whole system could be
derived from ${\cal Z}^{(0)}[{\bf j}, {\bf \bar j}]= 
\Pi_i {\cal Z}_i^{(0)}[{\bf j}_i, {\bf \bar j}_i]$. Whereas an expansion of each
${\cal
Z}^{(0)}_i[{\bf j}_i, {\bf \bar j}_i]$ 
in the currents would be 
completely local, the expansion of ${\cal Z}^{(0)}[{\bf j}, {\bf \bar j}]$ 
would contain highly non-local objects as well. Nevertheless it 
is clear, that these non-local Green functions arise as products of local
Green functions, and it must therefore be 
possible to find a decomposition. This mixing is circumvented from the
beginning, if we consider 
instead ${\cal F}^{(0)}[{\bf j}, {\bf \bar j}] 
= -i \sum_i \ln{\cal Z}_i^{(0)}[{\bf j}_i, {\bf \bar j}_i]$, 
where the logarithm has turned the product of subsystems into a sum of them.

In the system with hopping, of course, the situation is not that simple, but if
we expand ${\cal F}[{\bf j}, {\bf \bar j}]$ not only in the currents, but also
in the
hopping parameter $J$, the expansion still contains only the local cumulants of
the unperturbed system, and the hopping simply appears as a link between two
cumulants. For an expansion up to fourth order in the currents, we will need the
unperturbed cumulants $C_i^{\mathrm{K}_1\mathrm{K}_2(0)} (t_1;t_2)$ and
$C_i^{\mathrm{K}_1\mathrm{K}_2\mathrm{K}_3\mathrm{K}_4(0)}(t_1,t_2;t_3,t_4)$,
where $\mathrm{K}=\Sigma,\Delta$ denotes the respective Keldysh index. All the
cumulants
have only  one site index due to their locality. Note that cumulants describing
an unequal number of annihilation and creation processes vanish in the
unperturbed system, since $\op{H}_0$ commutes with the local number operator
$\op{n}_i \equiv \op{a}_i^\dagger \op{a}_i$. Their relations to the unperturbed
Green functions can be
derived from Eqs.~(\ref{GD}) and (\ref{CD}) and read explicitly
\begin{align}
\label{deco}
 C_i^{\mathrm{K}_1\mathrm{K}_2(0)} (t_1; t_2) =
G^{\mathrm{K}_1\mathrm{K}_2(0)}_{ii} (t_1;t_2),
\end{align}
and
\begin{align}
 \label{decomp}
&  C_i^{\mathrm{K}_1\mathrm{K}_2\mathrm{K}_3\mathrm{K}_4(0)}(t_1,t_2;t_3,t_4) 
 \\ & = G^{\mathrm{K}_1\mathrm{K}_2\mathrm{K}_3\mathrm{K}_4(0)}_{iiii}
(t_1,t_2;t_3,t_4) 
 - i C_i^{\mathrm{K}_1\mathrm{K}_3(0)} (t_1; t_3) \nonumber\\ & \nonumber
 \times C_i^{\mathrm{K}_2\mathrm{K}_4(0)} (t_2; t_4)-
i C_i^{\mathrm{K}_1\mathrm{K}_4(0)} (t_1; t_4) C_i^{\mathrm{K}_2\mathrm{K}_3(0)}
(t_2; t_3).
\end{align}
The unperturbed Green  functions are obtained by setting $J=0$ in
Eq.~(\ref{HD}). Explicitly, they read:
\begin{align}
 G^{\mathrm{K}_1\mathrm{K}_2(0)}_{ij} (t_1;t_2) = \frac{ \mathrm{Tr} \left\{
\mathrm{e}^{-\beta \op{H}_0} \op{T}_c \left[
\op{a}_i^{\mathrm{K}_1}(t_1)\op{a}_j^{\mathrm{K}_2\dagger}(t_2)
\right] \right\}}{\mathrm{Tr} \left\{ \mathrm{e}^{-\beta \op{H}_0} \right\}},
\end{align}
and
\begin{align}
& G^{\mathrm{K}_1\mathrm{K}_2\mathrm{K}_3\mathrm{K}_4(0)}_{ijkl}
(t_1,t_2;t_3,t_4) = 
\\ & \nonumber \frac{ \mathrm{Tr} \left\{ \mathrm{e}^{-\beta \op{H}_0} \op{T}_c
\left[ \op{a}_i^{\mathrm{K}_1}(t_1)\op{a}_j^{\mathrm{K}_2}(t_2)
\op{a}_k^{\mathrm{K}_3\dagger}(t_3)\op{a}_l^{\mathrm{K}_4\dagger}
(t_4)\right]\right\} }{\mathrm{Tr} \left\{ \mathrm{e}^{-\beta \op{H}_0}\right\}
}.
\end{align}
As argued at the end of Section \ref{keldysh}, we have $G^{\Delta\Delta} = C^{\Delta\Delta}
=
G^{\Delta\Delta\Delta\Delta} = C^{\Delta\Delta\Delta\Delta} = 0$, from which
follows that
$C_i^{\mathrm{K}_1\mathrm{K}_2\mathrm{K}_3\mathrm{K}_4}=
\delta_{ij} \delta_{jk} \delta_{kl} G_{ijkl}^{\mathrm{K} _1\mathrm {
K}_2\mathrm{K}_3\mathrm{K}_4}$, if three of the four Keldysh indices 
$\mathrm{K} _1, \mathrm {K}_2, \mathrm{K}_3, \mathrm{K}_4$ are $\Delta$. 

It is feasible to interpret the $2n$-point cumulants as tensors of rank
$2n$ in a two-dimensional space, which accounts for the doubled time degree of
freedom. We denote these tensors by
${\bf C}_i(t_1,\cdots,t_n;t_{n+1},\cdots,t_{2n})$. The most important objects
within the Schwinger-Keldysh formalism are the two-point cumulants, which form a
2x2 matrix:
\begin{align}
{\bf
C}_i(t_1,;t_2)
\equiv  \left( \begin{array}{cc} 
C_{i}^{\Sigma\Sigma(0)}(t_1,t_2) & C_{i}^{\Sigma\Delta(0)}(t_1,t_2)  \\
C_{i}^{\Delta\Sigma(0)}(t_1,t_2) & 0  \\
\end{array} \right)\, .
\end{align}
A comparison with Eq.~(\ref{krot}) together with Eq.~(\ref{deco}) yields that
the upper (lower) 
off-diagonal element are the retarded (advanced) Green functions
$G_{ii}^{\mathrm{R/A}}(t_1;t_2)$.

With the definition of vector operators and vector currents as in 
Eq.~(\ref{F1}) and the definition of tensor cumulants, we can formally
circumvent
using explicitly path or Keldysh indices. Within a diagrammatic notation this
simplifies the expansion of 
${\cal F}[{\bf j}, {\bf \bar j}]$, as developed for the Hubbard model in
Ref.~\cite{metzner}  and,
more recently, for the
Bose-Hubbard model within an ITF in Ref.~\cite{barry}. 
With minor modifications, these diagrammatic rules can be taken over to the 
Schwinger-Keldysh formulation as follows:
\begin{enumerate}
\item The building blocks of the expansion are the unperturbed $2n$-point
cumulants. They are represented by a black circle located at site $i$ with $n$
in-going and $n$ out-going legs. These legs carry the corresponding time
variables. In-going legs are associated with a particle creation process, while
out-going legs represent a particle annihilation process.
For example we have for $n=1$:
\begin{fmffile}{cumulant}
\begin{align}
\label{cumgraph}
\setlength{\unitlength}{1mm}
\parbox{10mm}{\begin{center}
\begin{fmfgraph*}(8,4) 
\setval 
\fmfstraight 
\fmfleft{i1}
\fmfright{o1}
\fmf{fermion}{i1,v1,o1}
\fmfv{decor.shape=circle,decor.full=full,decor.size=3.thick, label=${i}$,
l.dist=2mm, l.angle=90}{v1}
\fmfv{decor.size=0, label=${\scriptstyle t_2}$, l.dist=1mm, l.angle=90}{i1}
\fmfv{decor.size=0, label=${\scriptstyle t_1}$, l.dist=1mm, l.angle=90}{o1}
\end{fmfgraph*}
\end{center}} 
 = {\bf C}_{i} (t_1; t_2).
\end{align}
\item A hopping process from site $i$ to $j$ is described by linking one
out-going leg of the cumulant at site $i$ to an in-going leg of the cumulant
sitting on the neighboring site $j$. We have to associate these internal lines
with a factor $J$. The number of internal lines within one diagram constitutes
its hopping order. The time variable of an internal line has to be integrated.
A subtlety  coming from the CTPF in Keldysh space is the $\sigma^1$ matrix
between the vector operators in Eq.~(\ref{F1}), which defines the matrix
structure
of the internal lines \cite{rammer}.
\item Finally, the remaining legs have to be closed by a current, which are
represented by a black square:
\begin{align}
\label{curgraph}
\setlength{\unitlength}{1mm}
\hspace{5mm}
\parbox{8mm}{\begin{center}
\begin{fmfgraph*}(4,2) 
\setval 
\fmfstraight
\fmfleft{i1}
\fmfright{o1}
\fmf{fermion}{i1,o1}
\fmfv{decor.shape=square,decor.full=full,decor.size=3.thick}{i1}
\fmfv{decor.size=0, label=${\scriptstyle i}{\scriptstyle t}$, l.dist=1mm,
l.angle=90}{o1}
\end{fmfgraph*}
\end{center}}
= {\bf j}_i(t)
\parbox{12mm}{\begin{center}\rm{,}\end{center}}
\parbox{8mm}{\begin{center}
\begin{fmfgraph*}(4,2) 
\setval
\fmfstraight
\fmfleft{i1}
\fmfright{o1}
\fmf{fermion}{i1,o1}
\fmfv{decor.shape=square,decor.full=full,decor.size=3.thick}{o1}
\fmfv{decor.size=0, label=${\scriptstyle i}{\scriptstyle t}$, l.dist=1mm,
l.angle=90}{i1}
\end{fmfgraph*}
\end{center}}
={\bf \bar j}_i(t).
\end{align}
The distinction between currents ${\bf j}$ and conjugate currents ${\bf \bar
j}$ is taken care of by the direction of the line, since ${\bf j}$
(${\bf \bar j}$) is related to creation (annihilation) processes. Time
and space variables of the currents have to agree with the ones of the
corresponding leg and cumulant. 
\item In the free energy, all variables of any diagram have to be summed or
integrated, so we can save space by completely suppressing the variables as well
as the sums and integrals. For instance, the following closed graph has
to be interpreted as
\begin{align}
&
\label{simdia}
\setlength{\unitlength}{1mm}
\hspace{5mm}
\parbox{25mm}{\begin{center}
\begin{fmfgraph*}(18,8) 
\fmfkeep{simdia}
\setval
\fmfstraight
\fmfleft{i1}
\fmfright{o1}
\fmf{fermion}{i1,v1,o1}
\fmfv{decor.shape=square,decor.full=full,decor.size=3.thick}{i1}
\fmfv{decor.shape=square,decor.full=full,decor.size=3.thick}{o1}
\fmfv{decor.shape=circle,decor.full=full,decor.size=3.thick}{v1}
\end{fmfgraph*}
\end{center}} 
 = \sum_{ij}
 \int_{-\infty}^{\infty} \mathrm{d}t_1 \int_{-\infty}^{\infty} \mathrm{d}t_2 
\nonumber \\ &
\hspace{7mm}
\times \delta_{ij} {\bf \bar j}_{i}(t_1){\bf C}_i(t_1, t_2)
 {\bf j}_j(t_2).
\end{align}
\item Up to a given order in the hopping and in the currents, the free energy is
found by writing down all topologically 
inequivalent linked diagrams. In this diagrammatic notation, the statement of
the linked-cluster theorem becomes literal: 
Diagrams of distinct cumulants \textit{not} being linked via a hopping process
do \textit{not} contribute. The only 
remaining subtlety is the weight of each contributing diagram: On the one hand, we get a
factor $1/(m_j! m_j! m_J!)$ in the $m_J$th 
hopping order and $2m_j$th order in the currents from the Taylor expansion of
(\ref{F}). On the other hand, within one 
diagram there are exactly $m_j! m_j! m_J!$ lines, which can be interchanged. If
each permutation would
lead to a different term 
within the cumulant decomposition, both factors would cancel. But as we will
explicitly see in the fourth-order terms 
below, interchanging two lines or two vertices, might give exactly the same
diagram again. To avoid an over-counting of 
these diagrams we have to divide them by their respective
symmetry factor, i.e. by the number $M$ of possible permutations yielding the 
same diagram.
\end{fmffile}
\end{enumerate}

Setting $\hbar \equiv 1$ from now on, the whole expansion up to first hopping
order and fourth order in the current 
reads:
\begin{fmffile}{free-en}
\begin{align}
\label{F-dia}
& {\cal F}^{(4,1)}[{\bf j,\bar j}] =
\
\parbox{12mm}
{
\begin{center}
\begin{fmfgraph*}(12,6) 
\setval
\fmfstraight
\fmfleft{i1}
\fmfright{o1}
\fmf{fermion}{i1,v1,o1}
\fmfv{decor.shape=square,decor.size=3.thick}{i1}
\fmfv{decor.shape=square,decor.size=3.thick}{o1}
\fmfv{decor.shape=circle,decor.size=3.thick}{v1}
\end{fmfgraph*}
\end{center}
}
\ \ + \ \
 \parbox{15mm}{\begin{center}
\begin{fmfgraph*}(15,6)
\fmfkeep{hopdia2}
\setval
\fmfstraight
\fmfleft{i1}
\fmfright{o1}
\fmf{fermion}{i1,v1,v2,o1}
\fmfv{decor.shape=square,decor.size=3.thick}{i1}
\fmfv{decor.shape=square,decor.size=3.thick}{o1} 
\fmfv{decor.shape=circle,decor.size=3.thick}{v1}
\fmfv{decor.shape=circle,decor.size=3.thick}{v2}
\end{fmfgraph*}
\end{center}}
\\
\nonumber
&  
\hspace*{-2mm}+ \frac{1}{4} \hspace*{1mm}\parbox{16mm}{\begin{center}
\begin{fmfgraph*}(15,8)
\setval
\fmfstraight
\fmfkeep{vertex}
\fmfleft{i1,i2}
\fmfright{o1,o2}
\fmf{fermion}{i1,v1,o2}
\fmf{fermion}{i2,v1,o1}
\fmfv{decor.shape=square,decor.filled=full,decor.size=3.thick}{i1}
\fmfv{decor.shape=square,decor.filled=full,decor.size=3.thick}{i2}
\fmfv{decor.shape=square,decor.filled=full,decor.size=3.thick}{o1}
\fmfv{decor.shape=square,decor.filled=full,decor.size=3.thick}{o2}
\fmfv{decor.shape=circle,decor.filled=full,decor.size=3.thick}{v1}
\end{fmfgraph*}
\end{center}}
\
\hspace*{-1mm}+\frac{1}{2} \Bigg[ \ \ 
\parbox{19mm}{\begin{center}
\begin{fmfgraph*}(18,8) 
\setval
\fmfstraight
\fmfkeep{vertex-rein}
\fmfleft{i1,i2}
\fmfright{o1,o2}
\fmf{fermion}{i1,v1,o2} 
\fmf{fermion}{i2,v2,v1,o1}
\fmfv{decor.shape=square,decor.filled=full,decor.size=3.thick}{i1}
\fmfv{decor.shape=square,decor.filled=full,decor.size=3.thick}{i2}
\fmfv{decor.shape=square,decor.filled=full,decor.size=3.thick}{o1}
\fmfv{decor.shape=square,decor.filled=full,decor.size=3.thick}{o2}
\fmfv{decor.shape=circle,decor.filled=full,decor.size=3.thick}{v1}
\fmfv{decor.shape=circle,decor.filled=full,decor.size=3.thick}{v2}
\end{fmfgraph*}
\end{center}
}
 \ + \
\parbox{19mm}{\begin{center}
\begin{fmfgraph*}(18,8)
\setval
\fmfstraight
\fmfkeep{vertex-raus}
\fmfleft{i1,i2}
\fmfright{o1,o2}
\fmf{fermion}{i1,v1,o2}
\fmf{fermion}{i2,v1,v2,o1}
\fmfv{decor.shape=square,decor.filled=full,decor.size=3.thick}{i1}
\fmfv{decor.shape=square,decor.filled=full,decor.size=3.thick}{i2}
\fmfv{decor.shape=square,decor.filled=full,decor.size=3.thick}{o1}
\fmfv{decor.shape=square,decor.filled=full,decor.size=3.thick}{o2}
\fmfv{decor.shape=circle,decor.filled=full,decor.size=3.thick}{v1}
\fmfv{decor.shape=circle,decor.filled=full,decor.size=3.thick}{v2}
\end{fmfgraph*} 
\end{center}
}
\ \ \ \Bigg].
\end{align}
\end{fmffile}
As already mentioned, we will obtain a Ginzburg-Landau functional by performing
a Legendre transformation. We 
therefore note that the order parameter fields and the currents are conjugate
variables:
\begin{align}
\label{Psi}
\frac{\delta {\cal F}[{\bf j,\bar j}]}{\delta {\bf \bar
j}_i(t)} = \big\langle \op{\bf a}_i(t)
\big\rangle_{\mathrm{H}_0} \equiv {\bf \Psi}_i(t) 
\end{align}
It might be astonishing that also the physical observables 
\begin{align}
{\bf \Psi}_i(t) 
= 
\frac{1}{\sqrt{2}}\left(
\begin{array}{c}
 \Psi_i^+(t) + \Psi_i^-(t) \\
 \Psi_i^+(t) - \Psi_i^-(t)
\end{array}
\right)\, ,
\end{align}
which naturally cannot take different values 
at the same time, appear as contour-ordered quantities.
However, we will later find equations of motion which 
determine the order parameter fields $\Psi_i^{\mathrm{P}}(t)$
and we will see that the ansatz $\Psi_i^+ (t)= \Psi_i^-(t)$ yields 
the physical solution of these equations. Nevertheless, for the time being, we have to keep the path index
$\mathrm{P}$ during the
envisioned Legendre transformation. 
\section{Resummation via Legendre Transformation\label{resum}}
We now define the Legendre transformation in the standard way, yielding a
functional $\Gamma[{\bf \Psi,\bar \Psi}]$, that 
we will refer to as the \textit{effective action} of the system:
\begin{align}
\label{legendre}
& \Gamma[{\bf \Psi,\bar \Psi}] \equiv {\cal F}\left[{\bf j,\bar j}\right] - 
\sum_i \int_{-\infty}^{\infty} \mathrm{d}t \big[ {\bf \bar j}_i(t) \cdot {\bf
\Psi}_i(t) + \mathrm{c.c.} \big].
\end{align}
To obtain an expansion of $\Gamma$ as a power series in $J, {\bf \Psi}$ and
${\bf \bar \Psi}$, we first insert the expansion of ${\cal F}[{\bf j,\bar j}]$
from Eq.~(\ref{F-dia}) into the definition of ${\bf \Psi}$ in Eq.~(\ref{Psi}),
yielding an expression of the order parameter fields as a power series in $J$,
${\bf j}$, and ${\bf \bar j}$. In the diagrammatic notation, the derivative in
Eq.~(\ref{Psi}) is obtained by taking away a black square with an in-going leg
from the
graphs in ${\cal F}$ and, if there are more than one such squares within one
graph, by applying the usual product 
rule of differentiation. With this we obtain for the
order parameter fields, which we denote by 
white squares, the following expression:
\begin{fmffile}{deriv}
\begin{align}
& \label{abl}
{\bf \Psi}_i(t) \equiv
\setlength{\unitlength}{1mm}
\parbox{8mm}{\begin{center}
\begin{fmfgraph*}(4,4) 
\setval 
\fmfstraight
\fmfleft{i1}
\fmfright{o1}
\fmf{fermion}{i1,o1}
\fmfv{decor.shape=square,decor.filled=empty,decor.size=3.thick}{i1}
\fmfv{decor.size=0, label=${\scriptstyle i}{\scriptstyle t}$, l.dist=1mm,
l.angle=90}{o1}
\end{fmfgraph*}
\end{center}} 
\equiv
\parbox{12.5mm}{\begin{center}
\begin{fmfgraph*}(9,4) 
\setval
\fmfstraight
\fmfleft{i1}
\fmfright{o1}
\fmf{fermion}{i1,v1,o1}
\fmfv{decor.shape=square,decor.size=3.thick}{i1}
\fmfv{ label=${\scriptstyle t}$, l.dist=0.5mm, l.angle=90}{o1}
\fmfv{decor.shape=circle,decor.size=3.thick, label=${\scriptstyle i}$,
l.dist=2mm, l.angle=90}{v1}
\end{fmfgraph*}
\end{center}} + 
\parbox{15mm}{\begin{center}
\begin{fmfgraph*}(13,4) 
\setval
\fmfstraight
\fmfleft{i1}
\fmfright{o1}
\fmf{fermion}{i1,v1,v2,o1} 
\fmfv{decor.shape=square,decor.size=3.thick}{i1}
\fmfv{label=${\scriptstyle t}$, l.dist=0.5mm, l.angle=90}{o1}
\fmfv{decor.shape=circle}{v2}
\fmfv{decor.shape=circle,decor.size=3.thick}{v1} 
\fmfv{decor.shape=circle,decor.size=3.thick, label=${\scriptstyle i}$,
l.dist=2mm, l.angle=90}{v2}
\end{fmfgraph*}
\end{center}}
+
\frac{1}{2}
\parbox{17.5mm}{\begin{center}
\begin{fmfgraph*}(15,5)
\setval
\fmfstraight
\fmfleft{i1,i2}
\fmfright{o1,o2}
\fmf{fermion}{i1,v1,o2}
\fmf{fermion}{i2,v1,o1}
\fmfv{decor.shape=square,decor.filled=full,decor.size=3.thick}{i1}
\fmfv{decor.shape=square,decor.filled=full,decor.size=3.thick}{i2}
\fmfv{label=${\scriptstyle t}$, l.dist=1mm, l.angle=-90}{o1}
\fmfv{decor.shape=square,decor.filled=full,decor.size=3.thick}{o2}
\fmfv{decor.shape=circle,decor.filled=full,decor.size=3.thick,label=${
\scriptstyle i}$, l.dist=2mm, l.angle=90}{v1}
\end{fmfgraph*}
\end{center}}
\nonumber
\\
& \hspace*{-2mm}+\frac{1}{2} \Bigg[2\,
\parbox{20mm}{\begin{center}
\begin{fmfgraph*}(18,5) 
\setval
\fmfstraight
\fmfleft{i1,i2}
\fmfright{o1,o2}
\fmf{fermion}{i1,v1,o2} 
\fmf{fermion}{i2,v2,v1,o1}
\fmfv{decor.shape=square,decor.filled=full,decor.size=3.thick}{i1}
\fmfv{decor.shape=square,decor.filled=full,decor.size=3.thick}{i2}
\fmfv{label=${\scriptstyle t}$, l.dist=1mm, l.angle=-90}{o1}
\fmfv{decor.shape=square,decor.filled=full,decor.size=3.thick}{o2}
\fmfv{decor.shape=circle,decor.filled=full,decor.size=3.thick,label=${
\scriptstyle i}$, l.dist=2mm, l.angle=90}{v1}
\fmfv{decor.shape=circle,decor.filled=full,decor.size=3.thick}{v2}
\end{fmfgraph*}
\end{center}
}
+
\parbox{20mm}{\begin{center}
\begin{fmfgraph*}(18,5)
\setval
\fmfstraight
\fmfleft{i1,i2}
\fmfright{o1,o2}
\fmf{fermion}{i1,v1,o2}
\fmf{fermion}{i2,v1,v2,o1}
\fmfv{decor.shape=square,decor.filled=full,decor.size=3.thick}{i1}
\fmfv{decor.shape=square,decor.filled=full,decor.size=3.thick}{i2}
\fmfv{label=${\scriptstyle t}$, l.dist=1mm, l.angle=-90}{o1}
\fmfv{decor.shape=square,decor.filled=full,decor.size=3.thick}{o2}
\fmfv{decor.shape=circle,decor.filled=full,decor.size=3.thick}{v1}
\fmfv{decor.shape=circle,decor.filled=full,decor.size=3.thick,label=${
\scriptstyle i}$, l.dist=2mm, l.angle=-90}{v2}
\end{fmfgraph*}  
\end{center}
}
+ 
\parbox{20mm}{\begin{center}
\begin{fmfgraph*}(18,5)
\setval
\fmfstraight
\fmfleft{i1,i2}
\fmfright{o1,o2}
\fmf{fermion}{i1,v1,o2}
\fmf{fermion}{i2,v1,v2,o1}
\fmfv{decor.shape=square,decor.filled=full,decor.size=3.thick}{i1}
\fmfv{decor.shape=square,decor.filled=full,decor.size=3.thick}{i2}
\fmfv{decor.shape=square,decor.filled=full,decor.size=3.thick}{o1}
\fmfv{label=${\scriptstyle t}$, l.dist=1mm, l.angle=90}{o2}
\fmfv{decor.shape=circle,decor.filled=full,decor.size=3.thick,label=${
\scriptstyle i}$, l.dist=2mm, l.angle=90}{v1}
\fmfv{decor.shape=circle,decor.filled=full,decor.size=3.thick}{v2}
\end{fmfgraph*} 
\end{center}
}
\Bigg].
\end{align}
\end{fmffile}
Variables fixed by the derivative appear explicitly in the diagrams, for all
other variables the 
previous summation convention continues to hold.

Inverting expression (\ref{abl})
iteratively in $J$ and ${\bf j}$, we find a power series of ${\bf j}[{\bf
\Psi,\bar \Psi}](J)$. 
In zeroth hopping order and first order in the currents, only the first diagram
of Eq.~(\ref{abl}) must be taken into 
account. Thus, the inversion reads:
\begin{fmffile}{inv}
\begin{align}
\label{j0}
\setlength{\unitlength}{1mm}
\parbox{7.5mm}{\begin{center}
\begin{fmfgraph*}(4,4) 
\setval 
\fmfstraight
\fmfleft{i1}
\fmfright{o1}
\fmf{fermion}{i1,o1}
\fmfv{decor.shape=square,decor.filled=full,decor.size=3.thick}{i1}
\fmfv{decor.size=0, label=${\scriptstyle i}{\scriptstyle t}$, l.dist=1mm,
l.angle=90}{o1}
\end{fmfgraph*}
\end{center}} = 
\parbox{12.5mm}{\begin{center}
\begin{fmfgraph*}(9,4) 
\setval
\fmfstraight 
\fmfkeep{h1}
\fmfleft{i1}
\fmfright{o1}
\fmf{fermion}{i1,v1,o1}
\fmfv{decor.shape=square,decor.filled=empty,decor.size=3.thick}{i1}
\fmfv{ label=${\scriptstyle t}$, l.dist=0.5mm, l.angle=90}{o1}
\fmfv{decor.shape=circle,decor.filled=empty,decor.size=3.thick,
label=${\scriptstyle i}$, l.dist=2mm, l.angle=90}{v1}
\end{fmfgraph*}
\end{center}} \, ,
\end{align}
where the white circle stands for the inverse cumulant $\left[{\bf
C}_i(t_1;t_2)\right]^{-1}$. Note that the triangular 
structure of ${\bf C}_i(t_1;t_2)$ is conserved under inversion:
\begin{align}
& \left[{\bf
C}_i(t_1;t_2)\right]^{-1} = \\ \nonumber &
= \left( \begin{array}{cc} 
0  & \left[C_{i}^{\Delta\Sigma(0)}(t_1;t_2)\right]^{-1}  \\
\left[C_{i}^{\Sigma\Delta(0)}(t_1;t_2)\right]^{-1} & \left[\tilde
C_{i}^{\Sigma\Sigma(0)}(t_1;t_2) \right]^{-1} \\
\end{array} \right),
\end{align}
where we have introduced the abbreviation
\begin{eqnarray}
&& \Big[\tilde C_{i}^{\Sigma\Sigma(0)}(t_1,t_2) \Big]^{-1} \equiv \\
&& \hspace*{-8mm}-C_{i}^{\Sigma\Sigma(0)}(t_1,t_2) 
\Big[C_{i}^{\Delta\Sigma(0)}(t_1,t_2)\Big]^{-1}
\Big[C_{i}^{\Delta\Sigma(0)}(t_1,t_2)\Big]^{-1}\, .\nonumber
\end{eqnarray}
Re-inserting this in the second diagram of Eq.~(\ref{abl}), we get the inversion
up to the first hopping order:
\begin{align}
\label{j1}
\setlength{\unitlength}{1mm}
\parbox{7.5mm}{\begin{center}
\begin{fmfgraph*}(4,4) 
\setval 
\fmfstraight
\fmfleft{i1}
\fmfright{o1}
\fmf{fermion}{i1,o1}
\fmfv{decor.shape=square,decor.filled=full,decor.size=3.thick}{i1}
\fmfv{decor.size=0, label=${\scriptstyle i}{\scriptstyle t}$, l.dist=1mm,
l.angle=90}{o1}
\end{fmfgraph*}
\end{center}} &= 
\parbox{12.5mm}{\begin{center}
\begin{fmfgraph*}(9,4) 
\setval
\fmfstraight 
\fmfkeep{h1}
\fmfleft{i1}
\fmfright{o1}
\fmf{fermion}{i1,v1,o1}
\fmfv{decor.shape=square,decor.filled=empty,decor.size=3.thick}{i1}
\fmfv{ label=${\scriptstyle t}$, l.dist=0.5mm, l.angle=90}{o1}
\fmfv{decor.shape=circle,decor.filled=empty,decor.size=3.thick,
label=${\scriptstyle i}$, l.dist=2mm, l.angle=90}{v1}
\end{fmfgraph*}
\end{center}}
- \sum_{j} J_{ij}
\parbox{7.5mm}{\begin{center}
\begin{fmfgraph*}(4,4) 
\setval 
\fmfstraight
\fmfkeep{h2}
\fmfleft{i1}
\fmfright{o1}
\fmf{fermion}{i1,o1}
\fmfv{decor.shape=square,decor.filled=empty,decor.size=3.thick}{i1}
\fmfv{decor.size=0, label=${\scriptstyle j}{\scriptstyle t}$, l.dist=1mm,
l.angle=90}{o1}
\end{fmfgraph*}
\end{center}}
\sigma^1
\end{align}
and, finally, taking into account the currents up to third order, we get
\begin{align}
\label{j}
& \setlength{\unitlength}{1mm}
\parbox{7.5mm}{\begin{center}
\begin{fmfgraph*}(4,4) 
\setval 
\fmfstraight
\fmfleft{i1}
\fmfright{o1}
\fmf{fermion}{i1,o1}
\fmfv{decor.shape=square,decor.filled=full,decor.size=3.thick}{i1}
\fmfv{decor.size=0, label=${\scriptstyle i}{\scriptstyle t}$, l.dist=1mm,
l.angle=90}{o1}
\end{fmfgraph*}
\end{center}} = 
\parbox{12.5mm}{\begin{center}
\begin{fmfgraph*}(9,4) 
\setval
\fmfstraight 
\fmfleft{i1}
\fmfright{o1}
\fmf{fermion}{i1,v1,o1}
\fmfv{decor.shape=square,decor.filled=empty,decor.size=3.thick}{i1}
\fmfv{ label=${\scriptstyle t}$, l.dist=0.5mm, l.angle=90}{o1}
\fmfv{decor.shape=circle,decor.filled=empty,decor.size=3.thick,
label=${\scriptstyle i}$, l.dist=2mm, l.angle=90}{v1}
\end{fmfgraph*}
\end{center}}
-
\frac{1}{2}
\parbox{21mm}{\begin{center}
\begin{fmfgraph*}(17,8)
\setval
\fmfstraight
\fmfleft{i1,i2}
\fmfright{o1,o2}
\fmf{fermion}{i1,v1,v2,v3,o2}
\fmf{fermion}{i2,v4,v2,v5,o1}
\fmfv{decor.shape=square,decor.filled=empty,decor.size=3.thick}{i1}
\fmfv{decor.shape=square,decor.filled=empty,decor.size=3.thick}{i2}
\fmfv{label=${\scriptstyle t}$, l.dist=1mm, l.angle=90}{o1}
\fmfv{decor.shape=square,decor.filled=empty,decor.size=3.thick}{o2}
\fmfv{decor.shape=circle,decor.filled=empty,decor.size=3.thick}{v1}
\fmfv{decor.shape=circle,decor.filled=empty,decor.size=3.thick}{v3}
\fmfv{decor.shape=circle,decor.filled=empty,decor.size=3.thick}{v4}
\fmfv{decor.shape=circle,decor.filled=empty,decor.size=3.thick,label=${
\scriptstyle i}$, l.dist=2mm, l.angle=-90}{v5}
\fmfv{decor.shape=circle,decor.filled=full,decor.size=3.thick}{v2}
\end{fmfgraph*}
\end{center}}
\nonumber \\ 
&- \sum_{j} J_{ij} \Bigg[
\parbox{7.5mm}{\begin{center}
\begin{fmfgraph*}(4,4) 
\setval 
\fmfstraight
\fmfleft{i1}
\fmfright{o1}
\fmf{fermion}{i1,o1}
\fmfv{decor.shape=square,decor.filled=empty,decor.size=3.thick}{i1}
\fmfv{decor.size=0, label=${\scriptstyle j}{\scriptstyle t}$, l.dist=1mm,
l.angle=90}{o1}
\end{fmfgraph*}
\end{center}}
+
\frac{1}{2} 
\parbox{21mm}{\begin{center}
\begin{fmfgraph*}(17,8)
\setval
\fmfstraight
\fmfleft{i1,i2}
\fmfright{o1,o2}
\fmf{fermion}{i1,v1,v2,v3,o2}
\fmf{fermion}{i2,v4,v2}
\fmf{fermion}{v2,o1}
\fmfv{decor.shape=square,decor.filled=empty,decor.size=3.thick}{i1}
\fmfv{decor.shape=square,decor.filled=empty,decor.size=3.thick}{i2}
\fmfv{label=${\scriptstyle t}$, l.dist=1.5mm, l.angle=90}{o1}
\fmfv{decor.shape=square,decor.filled=empty,decor.size=3.thick}{o2}
\fmfv{decor.shape=circle,decor.filled=empty,decor.size=3.thick}{v1}
\fmfv{decor.shape=circle,decor.filled=empty,decor.size=3.thick}{v3}
\fmfv{decor.shape=circle,decor.filled=empty,decor.size=3.thick}{v4}
\fmfv{label=${\scriptstyle j}$, l.dist=1.5mm, l.angle=90,
decor.shape=circle,decor.filled=full,decor.size=3.thick}{v2}
\end{fmfgraph*}
\end{center}}
\Bigg]\sigma^1.
\end{align}
\end{fmffile}
Note that the lines between a cumulant and its inverse are no internal lines,
i.e. they don't represent a hopping.
Inserting the result from Eq.~(\ref{j}) into Eq.~(\ref{legendre}) together with
(\ref{F-dia})
and discarding any diagram, which is higher than first 
order in the hopping or higher than fourth order in the order parameter
fields, we finally find the effective action in the shape of a 
Ginzburg-Landau functional in first hopping order: 
\begin{fmffile}{gamma}
\begin{align}
\label{gamSF}
& \Gamma^{(4,1)}[\Psi,\bar \Psi] = -
\parbox{15mm}{\begin{center}
\begin{fmfgraph*}(12,6) 
\setval
\fmfstraight 
\fmfleft{i1}
\fmfright{o1}
\fmf{fermion}{i1,v1,o1}
\fmfv{decor.shape=square,decor.filled=empty,decor.size=3.thick}{i1}
\fmfv{decor.shape=square,decor.filled=empty,decor.size=3.thick}{o1}
\fmfv{decor.shape=circle,decor.filled=empty,decor.size=3.thick}{v1}
\end{fmfgraph*}
\end{center}} 
\\  \nonumber &
+ \int_{- \infty}^\infty \mathrm{d}t \sum_{ij} \ \ 
\parbox{6mm}{\begin{center}
\begin{fmfgraph*}(6,6) 
\setval 
\fmfstraight
\fmfleft{i1}
\fmfright{o1}
\fmf{fermion}{i1,o1}
\fmfv{decor.shape=square,decor.filled=empty,decor.size=3.thick}{i1}
\fmfv{decor.size=0, label=${\scriptstyle i}{\scriptstyle t}$,
l.dist=1mm, l.angle=130}{o1}
\end{fmfgraph*}
\end{center}}
[J_{ij} \sigma^1]
\parbox{6mm}{\begin{center}
\begin{fmfgraph*}(6,6) 
\setval 
\fmfstraight
\fmfleft{i1}
\fmfright{o1}
\fmf{fermion}{i1,o1}
\fmfv{decor.shape=square,decor.filled=empty,decor.size=3.thick}{o1}
\fmfv{decor.size=0, label=${\scriptstyle j}{\scriptstyle t}$, l.dist=1mm,
l.angle=50}{i1}
\end{fmfgraph*}
\end{center}} \ 
+
\frac{1}{4} \ \ 
\parbox{20mm}{\begin{center}
\begin{fmfgraph*}(20,10)
\setval
\fmfstraight
\fmfleft{i1,i2}
\fmfright{o1,o2}
\fmf{fermion}{i1,v1,v2,v3,o2}
\fmf{fermion}{i2,v4,v2,v5,o1}
\fmfv{decor.shape=square,decor.filled=empty,decor.size=3.thick}{i1}
\fmfv{decor.shape=square,decor.filled=empty,decor.size=3.thick}{i2}
\fmfv{decor.shape=square,decor.filled=empty,decor.size=3.thick}{o1}
\fmfv{decor.shape=square,decor.filled=empty,decor.size=3.thick}{o2}
\fmfv{decor.shape=circle,decor.filled=empty,decor.size=3.thick}{v1}
\fmfv{decor.shape=circle,decor.filled=empty,decor.size=3.thick}{v3}
\fmfv{decor.shape=circle,decor.filled=empty,decor.size=3.thick}{v4}
\fmfv{decor.shape=circle,decor.filled=empty,decor.size=3.thick}{v5}
\fmfv{decor.shape=circle,decor.filled=full,decor.size=3.thick}{v2}
\end{fmfgraph*}
 \end{center}}.
\end{align}
\end{fmffile}
Comparing the graphical content of the effective action in Eq.~(\ref{gamSF})
with the corresponding one of the
free energy in Eq.~(\ref{F-dia}), we find that it has a much simpler structure:
In the last two diagrams of the free energy the hopping connects two cumulants,
yielding a so-called
one-particle reducible diagram, i.e. a diagram which might be divided into two
diagrams by cutting one single line. In the effective action, however, these
diagrams do not appear and only the one-particle irreducible diagrams remain
\cite{zinn-justin,schulte}.

At this place, it is helpful to express the result (\ref{gamSF}) 
in analytic terms. 
However, while in the diagrammatic approach, apart from a straight-forward
re-definition of the diagrams, it
makes no difference if we work in real space or in Fourier space, the analytic
expression is simplified considerably 
by a Fourier transformation. Therefore, we define the transformation into
frequency space
\begin{align}
\label{f-omega}
 f(\omega) = \int_{-\infty}^{\infty} \mathrm{d}t \ f(t) \mathrm{e}^{i \omega t},
\end{align}
and into wave vector space
\begin{align}
\label{f-k}
 f_{\vec{k}} = \sum_i f_i \mathrm{e}^{-i \vec{k} \cdot \vec{r}_i}.
\end{align}
For conjugate variables, e.g. ${\bf \bar \Psi}$, the corresponding conjugate
transformations
hold.

Since the unperturbed Hamiltonian $\op{H}_0$ in (\ref{H0}) is not explicitly
time-dependent,
the unperturbed cumulants may only depend on the differences in their time
variables. As we will explicitly see in the Appendix \ref{cum}, where all the
relevant
cumulants are calculated in frequency space, this independence from an absolute
time yields a Dirac $\delta$-function in frequency space. Thus we are able to
define
\begin{eqnarray}
  {\bf C}_i(\omega_1;\omega_2) &\equiv& \delta(\omega_1-\omega_2) {\bf
c}_i(\omega_1)
\end{eqnarray}
and we have
\begin{eqnarray}
  {\bf C}(\omega_1,\omega_2; \omega_3, \omega_4) &\sim&
\delta(\omega_1+\omega_2-\omega_3-\omega_4)\, .
\end{eqnarray}
Regarding the spatial variables, the homogeneity of $J_{ij}$ assures that the
hopping term
becomes local by a transformation into wave vector space. In a
cubic lattice with lattice spacing $a$, for instance, we get
\begin{align}
 J_{ij} \rightarrow 2 J \sum_{i=1}^3 \cos(k_i a) \equiv J_{\vec{k}}.
\end{align}
As all cumulants are local, the transformation into wave vector space always
yields a
$\vec{k}$-dependence in form of $\delta_{\vec{k},\vec{0}}$. Since also a
summation 
over $\vec{k}$ must be performed, we can completely suppress the wave vector
indices of the cumulants.

The effective action then reads in terms of Fourier transformed quantities:
\begin{widetext}
\begin{align}
\label{gamSF2}
 & \Gamma^{(4,1)}=
 -\sum_{\vec{k}} \int \mathrm{d}\omega \ {\bf \bar
\Psi}_{\vec{k}}(\omega) \Big\{
\left[{\bf c}(\omega)\right]^{-1}-
 \sigma_1 J_{\vec{k}} \Big\}
{\bf \Psi}_{\vec{k}}(\omega) 
+
\frac{1}{4} \sum_{\{\vec{k}\}} \int \mathrm{d}\{\omega \} \
\delta_{\vec{k}_1+\vec{k}_2-\vec{k}_3-\vec{k}_4}
\nonumber \\ & 
\times   {\bf \bar \Psi}_{\vec{k}_1}(\omega_1) \left[ {\bf
c}(\omega_1)\right]^{-1}
   {\bf \bar \Psi}_{\vec{k}_2}(\omega_2) \left[ {\bf c}(\omega_2)\right]^{-1}
{\bf C}(\omega_1,\omega_2;\omega_3,\omega_4)
   \left[ {\bf c}(\omega_3)\right]^{-1}{\bf \Psi}_{\vec{k}_3}(\omega_3)
   \left[ {\bf c}(\omega_4)\right]^{-1}{\bf \Psi}_{\vec{k}_4}(\omega_4).
\end{align}
\end{widetext}
Now we can show that this expression of the effective action contains
resummed 
hopping processes.
To this end, we consider the first term in Eq.~(\ref{gamSF2}), which is of
second order in the fields: 
$[{\bf c}(\omega)]^{-1} \left[ 1 - J_{\vec{k}} \sigma^1 {\bf c}(\omega)
\right]$. Expanding this expression into 
a geometric sum, yields
$\left( {\bf c}(\omega) \sum_n \left[ J_{\vec{k}} \sigma^1 {\bf c}(\omega)
\right]^n \right)^{-1}$.
Diagrammatically, this represents the inverse of a sum of arbitrarily long hopping
``chains''. Thus, by means of a resummation, any of these terms is
automatically taken into account within the effective action. This is the clue
which
transforms the initial perturbative hopping expansion for the free energy 
into an effective $1/D$ expansion for the effective action, from
which we may expect reliable results not only in the MI but also in the SF
phase.

We conclude this section by showing how to extract physical information from the
effective action given 
by Eq.~(\ref{gamSF2}). The easiest and most direct way to do so is by noting
that the physical situation 
corresponds to vanishing currents. Due to Legendre identities this means
\begin{eqnarray}
\label{eqm}
{\bf \bar j}_{\vec{k}}(\omega)=\frac{\delta \Gamma}{\delta {\bf
\Psi}_{\vec{k}}(\omega)} \stackrel{!}{=} {\bf 0},
\end{eqnarray}
and the complex conjugate of this equation. We will refer to these equations as
the \textit{equations of motion}
as they determine the order parameter field. 
In the next section we will deal with their solutions. This will yield both the
phase boundary, if we consider the 
static case, and the excitation spectra in the dynamical case.
\section{Equations of Motion \label{sem}}
Analyzing the structure of the effective action $\Gamma$ in (\ref{gamSF2}) we
find that the second-order term 
$\bar \Psi^\Sigma \Psi^\Sigma$ is related to the vanishing cumulant $C^{\Delta
\Delta}$ and the fourth-order term 
$\bar \Psi^\Sigma \bar \Psi^\Sigma \Psi^\Sigma \Psi^\Sigma$ is related to the
vanishing cumulant 
$C^{\Delta\Delta\Delta\Delta}$. This means that all non-vanishing terms in
Eq.~(\ref{gamSF2}) contain $\Psi^\Delta$ at 
least to first order. Since the first component of Eq.~(\ref{eqm}) is given by
$\delta \Gamma / \delta \Psi^{\Sigma}=0$, 
all the terms in this equation will contain $\Psi^\Delta$ or $\bar \Psi^\Delta$
as a factor. 
As we aim at solving the equations of motion with the ansatz $\Psi^+ (t)= \Psi^-
(t)$, i.e. with 
an order parameter field which is independent of the time path, we are looking for 
solutions $\Psi^\Delta = 0$. Thus, we find that the first 
component of Eq.~(\ref{eqm}) is always trivially fulfilled. The second component
reads:
\begin{widetext}
 \begin{align}
\label{eqm1}
& \frac{\delta \Gamma}{\delta \Psi_{\vec{k}}^\Delta(\omega)} = 
	-\bar \Psi_{\vec{k}}^{\Sigma}(\omega) \left(
\frac{1}{G^{\mathrm{A}(0)}(\omega)} - J_{\vec{k}} 
\right)
	+ \frac{1}{4}\int_{-\infty}^\infty \mathrm{d}\omega_1
\int_{-\infty}^\infty \mathrm{d}\omega_2 \int_{-\infty}^\infty 
\mathrm{d}\omega_3 \sum_{\vec{k}_1,\vec{k}_2,\vec{k}_3}
\delta_{\vec{k}_1+\vec{k}_2-\vec{k}_3-\vec{k}}
	 \nonumber \\ &
\times	\left[
C^{\Delta\Delta\Delta\Sigma(0)}(\omega_1,\omega_2;\omega_3,\omega)
	\ + \
C^{\Delta\Delta\Sigma\Delta(0)}(\omega_1,\omega_2;\omega,\omega_3)\right]
	\frac{\bar \Psi_{\vec{k}_1}^{\Sigma}(\omega_1)
	\bar \Psi_{\vec{k}_2}^{\Sigma}(\omega_2)
	\Psi_{\vec{k}_3}^{\Sigma}(\omega_3)}{G^{\mathrm{A}(0)}(\omega_1)
G^{\mathrm{A}(0)}(\omega_2) 
G^{\mathrm{R}(0)}(\omega_3) G^{\mathrm{A}(0)}(\omega)}
	\Bigg\} \stackrel{!}{=} 0.
\end{align}
\end{widetext}%

The four-point cumulants $C^{\mathrm{K_1 K_2 K_3
K_4}(0)}(\omega_1,\omega_2;\omega_3,\omega_4)$ are symmetric in $\{\omega_1, 
\mathrm{K_1}\} \leftrightarrow \{\omega_2,\mathrm{K_2}\}$ and  $\{\omega_3,
\mathrm{K_3}\} \leftrightarrow 
\{\omega_4,\mathrm{K_4}\}$, such that $
C^{\Delta\Delta\Delta\Sigma(0)}(\omega_1,\omega_2;\omega_3,\omega_4) 
=C^{\Delta\Delta\Sigma\Delta(0)}(\omega_1,\omega_2;\omega_4,\omega_3)$. Thus,
from the initially 16 different 4-point 
cumulants, only one has to be calculated. This is done in the appendix, where
also explicit expressions for the retarded 
and advanced Green functions $G^{\mathrm{R}(0)}$ and $G^{\mathrm{A}(0)}$ are
derived. There we will  see 
furthermore
that $2 C^{\Delta\Delta\Sigma\Delta(0)}(\omega_1,\omega_2,\omega_3,\omega_4)$
coincides with the advanced four-point
function $C^{\mathrm{A}(0)}(\omega_1,\omega_2;\omega_3,\omega_4)$ which is 
defined as the thermal average of multiple commutators 
times Heaviside step functions (cf. \cite{chou}). As the complex conjugate of
advanced Green functions are retarded Green
functions, the complex conjugate of Eq.~(\ref{eqm}) amounts to interchanging 
the respective retarded and advanced functions.

Due to the triple frequency integrals and the triple wave vector sums,
Eq.~(\ref{eqm1}) is not straightforwardly solved. 
In order to
continue analytically, we now assume the equilibrium ansatz
$\Psi^\Sigma_{\vec{k}}(\omega) = \delta(\omega) 
\delta_{\vec{k},\vec{0}} \Psi_{\mathrm{eq}}^\Sigma$ of a 
spatially homogeneous and temporally constant order parameter field. With that
ansatz, all integrals and sums become 
trivial, and we get the algebraic equation:
\begin{align}	
\label{e1}
	 \frac{1}{G^{\mathrm{A}(0)}(0)} - J_{\vec{0}} 
	=
	\frac{C^{\mathrm{A}(0)}(0,0;0,0) 
	\bar \Psi_{\mathrm{eq}}^\Sigma
	\Psi_{\mathrm{eq}}^\Sigma}{4G^{\mathrm{A}(0)}(0) G^{\mathrm{A}(0)}(0)
G^{\mathrm{R}(0)}(0) G^{\mathrm{A}(0)}(0)} \, .
\end{align}
Later on we will read off from this equation physically important results like
the phase boundary. At the 
moment we only need the information that 
this equation defines an equilibrium value $\Psi^\Sigma_{\mathrm{eq}}$ around
which we  
perform a harmonic approximation for the effective action
Eq.~(\ref{eqm1}) with the ansatz $\Psi_{\vec{k}}^{\mathrm{K}}(\omega) =
\Psi_{\mathrm{eq}}^{\mathrm{K}} 
+ \varPsi_{\vec{k}}^{\mathrm{K}}(\omega)$. This yields
\begin{widetext}
\begin{align}
\label{gex}
&\Gamma[\varPsi^{\Delta},\bar \varPsi^{\Delta},\varPsi^{\Sigma},\bar
\varPsi^{\Sigma}] \approx
\frac{1}{2} \sum_{\vec{k}_1,\vec{k}_2} \int_{-\infty}^{\infty}
\mathrm{d}\omega_1 \int_{-\infty}^{\infty} \mathrm{d}\omega_2 
 \Bigg\{ 
\frac{\delta^2 \Gamma}{\delta \Psi_{\vec{k}_1}^{\Sigma}(\omega_1) \delta \bar
\Psi_{\vec{k}_2}^{\Delta}(\omega_2)} 
\Bigg|_\mathrm{eq}
\bar
\varPsi_{\vec{k}_2}^{\Delta}(\omega_2)\varPsi_{\vec{k}_1}^{\Sigma}(\omega_1) 
+ \frac{\delta^2 \Gamma}{\delta \bar \Psi_{\vec{k}_1}^{\Sigma}(\omega_1) \delta
\Psi_{\vec{k_2}}^{\Delta}(\omega_2)} 
\Bigg|_\mathrm{eq}
 \\ & \nonumber
\times \varPsi_{\vec{k}_2}^{\Delta}(\omega_2)\bar
\varPsi_{\vec{k}_1}^{\Sigma}(\omega_1)
+ \frac{\delta^2 \Gamma}{\delta \Psi_{\vec{k}_1}^{\Sigma}(\omega_1) \delta
\Psi_{\vec{k}_2}^{\Delta}(\omega_2)} 
\Bigg|_\mathrm{eq}
\varPsi_{\vec{k}_2}^{\Delta}(\omega_2)\varPsi_{\vec{k}_1}^{\Sigma}(\omega_1) +
\frac{\delta^2 \Gamma}{\delta \bar\Psi_{\vec{k}_1}^{\Sigma}(\omega_1) \delta
\bar \Psi_{\vec{k}_2}^{\Delta}(\omega_2)} 
\Bigg|_\mathrm{eq}
 \bar \varPsi_{\vec{k}_2}^{\Delta}(\omega_2) \bar
\varPsi_{\vec{k}_1}^{\Sigma}(\omega_1) \Bigg\}.
\end{align}
\end{widetext}
Other derivatives either vanish or lead to terms which are of second order in
$\Psi^\Delta$. They would vanish 
when the resulting equations of motion are solved by $\Psi^\Delta=0$. Evaluating
explicitly
the respective derivatives in Eq.~(\ref{gex}), yields the linearized expression:
\begin{eqnarray}
&&\hspace*{-5mm}
\Bigg[
\frac{1}{G^{\mathrm{R}(0)}(\omega)}- J_{\vec{k}}
 - 
\frac{\left|\Psi_\mathrm{eq} \right|^2 C^{\mathrm{R}(0)}(\omega,0;0,\omega)}{2 G^{\mathrm{A}(0)}(0)
G^{\mathrm{R}(0)}(0) G^{\mathrm{R}(0)}(\omega) 
G^{\mathrm{R}}(\omega)}\Bigg] 
\nonumber \\ &&
\hspace*{-5mm}  \times \varPsi_{\vec{k}}^\Sigma(\omega)-
\frac{\left|\Psi_\mathrm{eq} \right|^2 C^{\mathrm{R}(0)}(\omega,-\omega;0,0)}{4 G^{\mathrm{A}(0)}(-\omega)
G^{\mathrm{R}(0)}(0) G^{\mathrm{R}(0)}(\omega) 
G_{\mathrm{R}(0)}(0)}
 \nonumber \\
&&\times \bar\varPsi^{\Sigma}_{-\vec{k}}(-\omega)\stackrel{!}{=}0 \, .
\label{em2}
\end{eqnarray}
Now we will investigate how this linearized equation of motion is related to the
Green functions. To this end
we start with the statement
\begin{align}
\label{2dev}
\left. \frac{\delta^2 {\cal F}}{\delta \bar j_i^{\Delta} \delta
j_j^{\Sigma}}\right|_{j=\bar j=0} = G_{ij}^{\mathrm{A}}\, .
\end{align}
Making use of the product rule for functional derivatives, we find the following
identity
\begin{eqnarray}
\label{dij}
\delta_{ij}&=& \frac{\delta j_{i}^{\Delta}}{\delta j_{j}^{\Delta}} = \sum_k \Bigg(
\frac{\delta j_{i}^{\Delta}}{\delta \Psi_{k}^{\Delta}} \frac{\delta
\Psi_{k}^{\Delta}}{\delta j_{j}^{\Delta}} 
+\frac{\delta j_{i}^{\Delta}}{\delta \bar \bar \Psi_{k}^{\Delta}} \frac{\delta
\bar \Psi_{k}^{\Delta}}{\delta j_{j}^{\Delta}}  \nonumber \\
&& +\frac{\delta j_{i}^{\Delta}}{\delta \Psi_{k}^{\Sigma}} \frac{\delta
\Psi_{k}^{\Sigma}}{\delta j_{j}^{\Delta}} 
+\frac{\delta j_{i}^{\Delta}}{\delta \bar \Psi_{k}{\Sigma}} \frac{\delta \bar
\Psi_{k}^{\Sigma}}{\delta j_{j}^{\Delta}} \Bigg),
\end{eqnarray}
and a similar one from $0 = \delta \bar j_{i}^{\Delta} / \delta j_{j}^{\Delta}$. With the Legendre identity
\begin{align}
 \Psi_{i}^{\Sigma} = \frac{\delta {\cal F}}{\delta \bar j_{i}^{\Delta}}\, ,
\end{align}
which is dual to (\ref{eqm}),
we are finally able to relate second derivatives of the effective action
$\Gamma$ to the advanced
two-point functions:
\begin{widetext}
\begin{align}
\label{GSF}
 G^{\mathrm{A}}_{\vec{k}}(\omega) = \frac{
\frac{\delta^2 \Gamma}{\delta \bar \Psi_{-\vec{k}}^{\Delta}(-\omega) \delta
\Psi_{-\vec{k}}^{\Sigma}(-\omega)}
\Big|_{\mathrm{eq}}
}{
\frac{\delta^2 \Gamma}{\delta \bar \Psi_{-\vec{k}}^{\Delta}(-\omega) \delta
\Psi_{-\vec{k}}^{\Sigma}(-\omega)}
\Big|_{\mathrm{eq}}
\frac{\delta^2 \Gamma}{\delta \Psi_{\vec{k}}^{\Delta}(\omega) \delta \bar
\Psi_{\vec{k}}^{\Sigma}(\omega)}\Big|_{\mathrm{eq}}-
\frac{\delta^2 \Gamma}{\delta \Psi_{-\vec{k}}^{\Delta}(-\omega) \delta
\Psi_{\vec{k}}^{\Sigma}(\omega)}\Big|_{\mathrm{eq}} 
\frac{\delta^2 \Gamma}{\delta \bar \Psi_{\vec{k}}^{\Delta}(\omega) \delta \bar
\Psi_{-\vec{k}}^{\Sigma}(-\omega)}
\Big|_{\mathrm{eq}}
}
\end{align}
\end{widetext}
and the complex conjugate expression for the retarded Green function.
Note that the equation of motion (\ref{em2}) is solved, when
the superfluid Green function (\ref{GSF}) diverges. However, the Green
functions contain more information than 
the equation of motion, since they allow, in principle,  
to extract also the respective spectral weights from its imaginary part as we will see below.
\section{Results \label{res}}
\begin{figure}[t]
\includegraphics[width=0.45\textwidth]{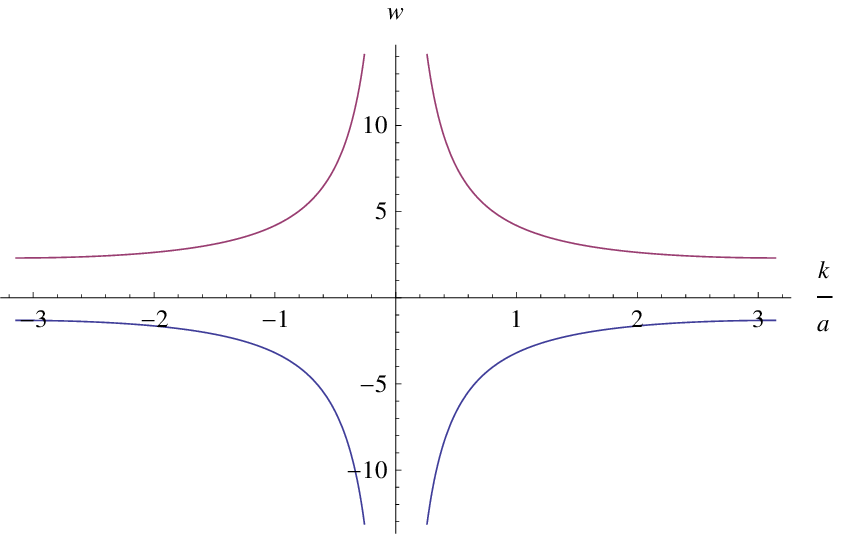}
\includegraphics[width=0.45\textwidth]{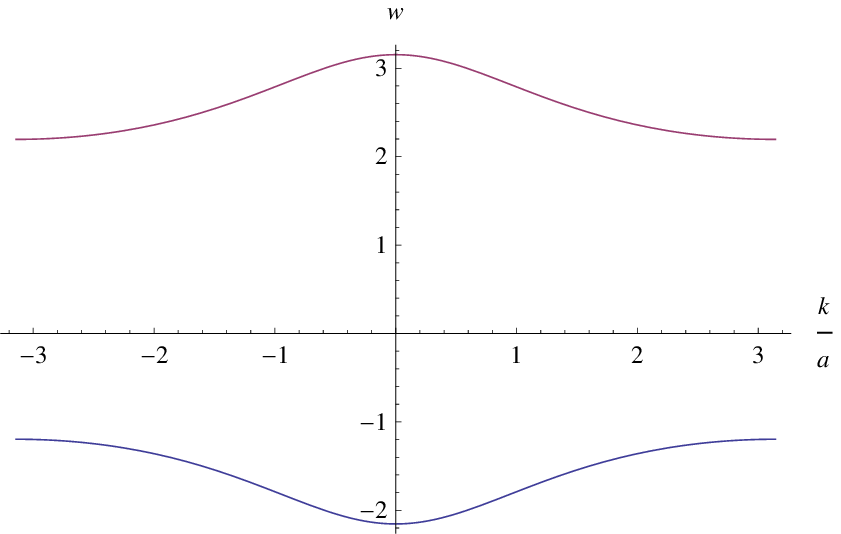}
\caption{\label{weights} Spectral weights at the lobe tip (up) and for
$J/U=0.02$
(below) as a function of $k=\vec{k} \cdot \vec{e}_x$.}
\end{figure}
\begin{figure*}
\includegraphics[width= 0.95 \textwidth]{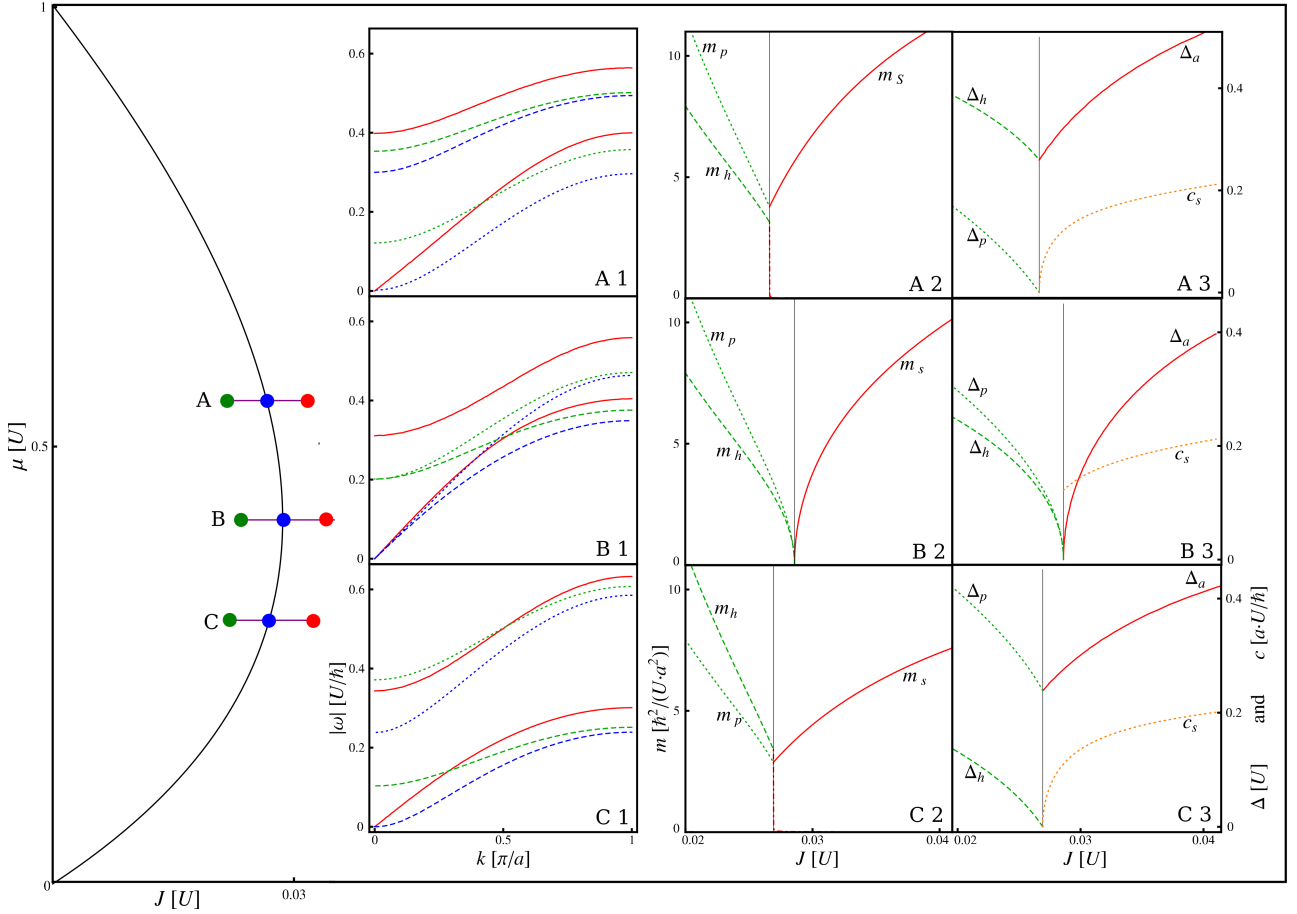}
\caption{\label{spectra} Excitations in $\boldsymbol{k}=(1,0,0)$ direction are
plotted in A1 -- C1 for different values 
$\mu/U$ and $J/U$, which are marked in the phase diagram (left). In the MI 
phase (green) and on the phase boundary (blue), the two $T=0$ modes can be
interpreted as particle (dotted) and hole 
(dashed) excitations. At the tip of the Mott lobe (B), both modes become
gapless, 
whereas for larger (smaller) chemical potentials $\mu$ only the gap of the
particle (hole) mode vanishes. In the SF 
phase (red), the gapless mode turns into a sound mode, but also a gapped mode
is 
present everywhere in the SF phase. The smooth transition from the MI excitation
to the SF excitation is further 
analyzed in A 2 -- C 2 and A 3 -- C 3, where the effective mass $m$ and the gap
$\Delta$ 
of each mode are plotted as a function of $J/U$. The sound velocity $c$ of the
massless SF excitation, plotted in 
A3 -- C3, vanishes at the phase boundary except at the tip, indicating the
existence 
of a different universality class in this configuration.}
\end{figure*}

The non-trivial solutions of Eq.~(\ref{eqm1}) can be divided into two classes
which we work out 
at $T=0$. How to perform the $T=0$ limit for the respective Green functions and
cumulants is shown in the appendix.

\subsection{Phase boundary}

Static solutions are obtained from Eq.~(\ref{e1}). The right-hand side of
Eq.~(\ref{e1}) does not explicitly depend on the 
hopping parameter, but it is multiplied by $|\Psi_{\mathrm{eq}}|^2 \geq 0$.
Thus, its sign is independent from $J$. The 
left-hand 
side of Eq.~(\ref{e1}) depends explicitly of $J$ and changes its sign at some
critical value $J_{\mathrm{PB}}(\mu,U)$, 
which is obtained by setting 
the left-hand side of Eq.~(\ref{e1}) equal to zero. From (\ref{EV}) and
(\ref{C2}) we get explicitly
\begin{align}
\label{pb}
 J_{\mathrm{PB}}(\mu,U)= -\frac{(U n-\mu-U)(U n- \mu)}{6(\mu/U + 1)}.
\end{align}
For $J<J_{\mathrm{PB}}$ at a given $\mu$ and $U$, only $\Psi_{\mathrm{eq}}=0$
solves Eq.~(\ref{e1}), and thus we are in 
the disordered, i.e. insulating phase. For $J>J_{\mathrm{PB}}$, the system is
superfluid.
A plot of the phase boundary (\ref{pb})
for $T=0$ and $0 \leq \mu/U \leq 1$ can be found on the left side of
Fig.~\ref{spectra} for $n=1$. 
In the 
considered first hopping order, this quantum phase diagram is identical to the
mean-field result \cite{stoof} with a 
deviation from recent high-precision Monte Carlo data \cite{montecarlo} of about
25~\%. 
Within a Landau expansion the second-order hopping contribution has recently
been calculated analytically in 
Ref.~\cite{ednilson} decreasing the error to less than 2~\%. 
A numerical evaluation of 
higher hopping orders has even been shown to converge to a quantum phase diagram
which is indistinguishable from the 
Monte-Carlo result \cite{eckardt,holthaus1,holthaus2}.

For obtaining dynamic solutions we have to consider the linearized equation of
motion Eq.~(\ref{em2}). Numerically, 
we obtain the dispersion relations shown in Fig. \ref{spectra}, where the MI-SF
transition can clearly be observed by a qualitative change in the 
excitations of the system. 

\subsection{MI spectra}

In the MI phase, we find the gapped particle/hole excitations from mean-field
theory \cite{stoof}. At the phase boundary, we have to distinguish the tip from
the rest of the lobe in accordance with the 
critical theory of the BH model \cite{fisher,sachdev}. While at the tip of the
$m$th lobe, i.e. at $\mu_m/U = \sqrt{m(m+1)}
-1$, both excitations become gapless and linear for small $|\boldsymbol{k}|$, 
off the tip at $\mu>\mu_m$ ($\mu<\mu_m$) only the particle (hole) mode becomes
gapless and remains with a finite effective
mass.

Since in the MI phase and at the phase boundary we have
$|\Psi_{\mathrm{eq}}|=0$, we are able to determine analytic solutions of the
equations of motion Eq. (\ref{em2}):
\begin{align}
\label{MIspectra}
\omega_{\pm}(\vec{k}) = \frac{1}{2\hbar} \Big[ & U(2n-1)-2\mu-J_{\vec{k}} \\
\nonumber & \pm \sqrt{
U^2 -2J_{\vec{k}} U (2n+1) +J_{\vec{k}}^2} \hspace{1mm} \Big].
\end{align}

The Green function in Eq. (\ref{GSF}) reduces in the MI phase to
\begin{align}
\label{gmi}
G^{\mathrm{R(MI)}}_{\vec{k}}(\omega) =
\left[\frac{1}{G^{\mathrm{R}(0)}(\omega)} - J_{\vec{k}} \right]^{-1}.
\end{align}
Comparing this with the general spectral representation of the retarded/advanced
Green
functions \cite{pathint}
\begin{align}
\label{specrep}
G^{\mathrm{R/A}}_{\vec{k}}(\omega) = \int_{-\infty}^{\infty} \mathrm{d}\omega'
\rho(\omega',\vec{k}) \left( \frac{{\cal P}}{\omega-\omega'} \pm i\pi
\delta(\omega-\omega') \right),
\end{align}
where ${\cal P}$ denotes the principal value of the integration across the
singularity and $\rho(\omega,\vec{k})$ denotes the spectral function, we can also obtain
information about the spectral weights of the excitations, i.e. how much
of the total excitation energy is stored in each excitation. To this end we have
to divide Eq. (\ref{gmi}) into its real and imaginary part, where the latter is
due to an infinitesimal $i \epsilon$-shift of the poles from the real axis into
the complex
plane (see Appendix A). By performing the limit $\epsilon \rightarrow 0$, we
find 
\begin{eqnarray}
\label{sf}
\hspace*{-4mm}\rho(\omega,\vec{k})= \delta (\omega-\omega_+(\vec{k}))
w_+(\vec{k}) + \delta(\omega-\omega_-(\vec{k})) w_-(\vec{k})
\end{eqnarray}
with the respective weights 
\begin{align}
\label{sw}
  w_{\pm}(\vec{k}) = \frac{1}{2} \left(1 \pm \frac{U(1+2n)-J_{\vec{k}}}
{\sqrt{U^2-2J_{\vec{k}} U(2n+1) + J_{\vec{k}}^2}} \right).
\end{align}
Note that they
do not depend on the chemical potential. At the tip of the lobe, both
weights diverge at $\vec{k}=\vec{0}$. We can check the spectral function (\ref{sf}), (\ref{sw}) 
by noting that it obeys the sum rule \cite{pathint}:
\begin{align}
\label{rho-norm}
\int_{-\infty}^{\infty} \mathrm{d} \omega \ \rho(\omega,\vec{k}) = 1.
\end{align}
The spectral weights are plotted for two different ratios $J/U$ in Fig.
\ref{weights}.

\subsection{SF spectra}

Turning into the SF phase, the gapless mode rapidly looses its mass and has to
be identified with the Goldstone mode which 
arises due to the broken $U(1)$ symmetry. Indeed, within the Ginzburg-Landau
theory it turns out in the limit $\vec{k}
\rightarrow \vec{0}$ and $\omega\rightarrow0$, that the excitation $\delta\Psi (\omega, \vec{k})$
stems from variations of the phase. 
Within a slave-boson approach it has even been shown in Ref.~\cite{blatter1} 
that, also for general wave vectors $\vec{k}$,
phase variations dominate this excitation. Thus density variations arise which
make this mode sensitive to Bragg 
spectroscopy. 
Recently, the whole sound mode has been measured via Bragg spectroscopy far away
from the 
phase boundary and could be well described via a Bogoliubov fit
\cite{sengstock}. Analytical results for this mode can be 
obtained by expanding Eq.~(\ref{em2}) for small $\omega$ and small $\vec{k}$ up
to second order. The resulting algebraic equation is
solved by the ansatz $\omega = c \, |\vec{k}|$. 
Defining the dimensionless quantities $\tilde \mu = \mu/U$ and $\tilde J = J/U$,
we find that the sound velocity $c$ 
of the Goldstone mode at $T=0$, which is given by the dimensionless quantity 
\begin{align}
 \tilde c(n,\tilde \mu,\tilde J) = \frac{c(n,\mu,J,U)}{aU/\hbar}
\end{align}
along any lattice direction. Here $n$ gives the occupation number of the MI
lobe, above which the SF theory is constructed.
For $n=1$, for instance, we find the explicit result
\begin{widetext}
\begin{align}
\label{cschlange}
 & \tilde c(1,\tilde \mu,\tilde J)  = 
\Big\{ \Big[ \tilde J \tilde\mu ^2 (1+\tilde \mu )^3 (3-2 \tilde\mu) \big(-3+8
\tilde\mu -10 \tilde\mu ^2+4 \tilde\mu ^3
+\tilde\mu ^4\big)^2 
\big((\tilde\mu -1) \tilde\mu +6 \tilde J (1+\tilde\mu )\big) \Big] \Big/
\nonumber \\ &
\Big[36 \tilde J (1-\tilde\mu )^3 \tilde\mu ^3 (-27+108 \tilde\mu +9 \tilde\mu
^2-92 \tilde\mu ^3+3 \tilde\mu ^4-24 
\tilde\mu ^5+7 \tilde\mu ^6)
-(-1+\tilde\mu )^3 \tilde\mu ^3 
 \\ &
(27-135 \tilde\mu +36 \tilde\mu ^2+172 \tilde\mu ^3-210 \tilde\mu ^4+294
\tilde\mu ^5-196 \tilde\mu ^6+60 \tilde\mu ^7
+15 \tilde\mu ^8+\tilde\mu ^9)
+18 \tilde J^2 (1+\tilde\mu )^2 
\nonumber \\ &
(27-270 \tilde\mu +1359 \tilde\mu ^2-3860 \tilde\mu ^3+5950 \tilde\mu ^4-4512
\tilde\mu ^5+1198 \tilde\mu ^6+100 
\tilde\mu ^7+135 \tilde\mu ^8-66 \tilde\mu ^9+3 \tilde\mu ^{10})\Big]\Big\}^{1/2}
\nonumber.
\end{align}
\end{widetext}
Additionally to that sound mode, however, also a gapped mode survives the
quantum phase transition or arises again if we 
depart from the lobe tip. It can be smoothly mapped onto one of the 
respective MI modes, which is shown in the plot of both the effective masses and
the gaps on the right side of 
Fig.~\ref{spectra}. The existence of such a SF gapped mode is in accordance with
theoretical results 
obtained previously in Refs.~\cite{dupuis,blatter1,menotti}, and it has also
recently been confirmed experimentally 
\cite{gapped-mode}. 
This mode is interpreted in Refs.~\cite{blatter1, gapped-mode} as an amplitude
excitation which corresponds to an 
exchange between condensed and non-condensed particles at constant overall
density. 
Although Eq.~(\ref{em2}) does not allow \textit{pure} amplitude excitations, we
can back this interpretation of 
predominant 
amplitude excitations by observing that the zero-momentum energy transfer at the
phase boundary corresponds to the 
creation of a particle/hole pair.

\begin{figure}[t]
\begin{centering}
\includegraphics[width= 0.45\textwidth]{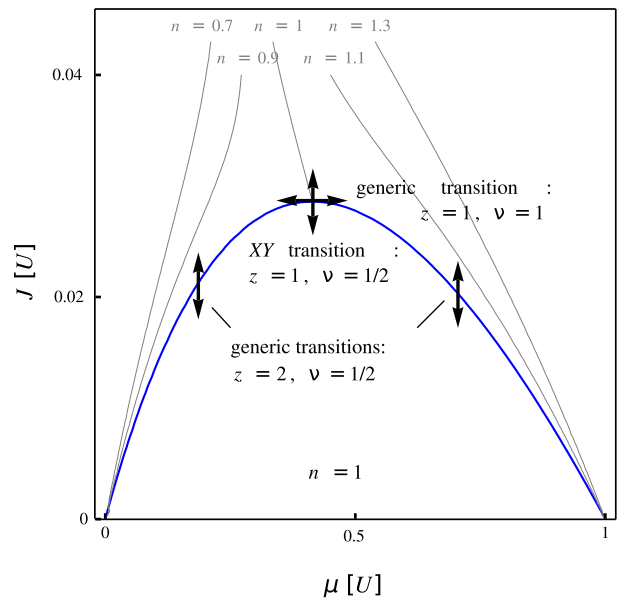}
\caption{\label{critex} The critical exponents depend on position and direction
of the phase transition. The grey lines 
of constant densities in the SF phase are obtained via a thermodynamic effective
action based on an analogue expansion in 
imaginary time \cite{barry}.}
\end{centering}
\end{figure}

\subsection{Critical exponents}

From our excitation spectra we can directly read off the dynamical critical exponent $z$,
which is $z=1$
at the tip, where a relativistic dispersion relation is found, and $z=2$ off the
tip, where the system behaves like a 
normal Bose gas. This is in agreement with the critical theory of the BH model
\cite{fisher,sachdev}. Physically, the 
difference between the ``generic'' transition off the lobe tip and the
``XY-like'' one at the tip is caused by the 
density variations that are only absent at the lobe tip. Writing the gap
$\Delta$ as a function of 
$J-J_{\rm PB}(\mu)$, i.e. 
of the distance to the phase boundary, we find that $\Delta \sim
[J-J_{\rm PB}(\mu)]^{z\nu}$ with the mean-field exponent 
$\nu=1/2$ for any \textit{vertical} phase transition, see Fig.~\ref{critex}.
Since vertical transitions really cross the 
phase boundary, their exponents can be obtained analytically by considering only
the behavior within the Mott phase. At 
the tip, there is also the possibility to touch the phase boundary horizontally,
i.e. without entering the MI phase. We 
find that this is a special case of the generic transitions with $z\nu=1$, where
$z=1$ and $\nu=1$ 
are obtained numerically.

\subsection{Gross-Pitaevskii limit}

Surprisingly, also the regime deep in the SF phase turns out to be accessible
with the Ginzburg-Landau theory. 
Expanding the Green functions in $U$, in the lowest non-trivial order they do
not depend on temperature and reduce to
\begin{align}
& G^{\mathrm{R/A}(0)}(\omega_1,\omega_2) = -2\pi \frac{1}{\mu +\omega }
\delta(\omega_1-\omega_2) +O(U) \\
& C^{\mathrm{R/A}(0)}(\omega_1,\omega_2;\omega_3,\omega_4) = -2\pi
\delta(\omega_1+\omega_2-\omega_3-\omega_4) \nonumber \\ 
&  \times \frac{2 U}{(\mu +\omega_1) (\mu +\omega_2) (\mu +\omega_3) (\mu
+\omega_4)} + O(U^2) .
\end{align}
Inserting this into Eq. (\ref{eqm1}) leads, after a Fourier transformation into
real time and space, to the equation of motion
\begin{align}
\label{GPEQ}
i \frac{\partial \Psi_i^\Sigma}{\partial t} = -\sum_j J_{ij} \Psi_j^\Sigma -\mu
\Psi_i^\Sigma + \frac{U}{2} \Psi_i^\Sigma |\Psi_i^\Sigma|^2,
\end{align}
which is the lattice version of the Gross-Pitaevskii (GP) equation
\cite{polkovnikov}. From this follows the Bogoliubov sound mode of a fully
condensed system \cite{stoof}, which is given by
\begin{align}
\label{bogspec}
\omega_{\vec{k}} &= \sqrt{\epsilon_{\vec{k}}^2 + 2U n \epsilon_{\vec{k}}},
\end{align}
where $\epsilon_{\vec{k}}= 6 J - J_{\vec{k}}$ denotes the free dispersion.
For small $|\vec{k}|$, this dispersion is linear with the sound velocity
\begin{align}
\label{bogc}
\frac{c}{a/\hbar}= \sqrt{2J (\mu +6J)},
\end{align}
which coincides with the limit $\tilde c U$ from Eq. (\ref{cschlange}) for $U \rightarrow
0$.
\\
Since in this limit the sound mode predicted by
our theory agrees perfectly with the one derived from the GP equation, it is interesting to
inspect also the gapped mode which is not predicted by the GP theory.
Thus, we cannot first expand the equation of motion for small $U$ and then
consider the solutions, since this would reduce our equation of motion to the
GP equation. Instead, we may investigate what happens to the solutions of our
\textit{full} equations of motion for small $U$. Numerically we find that the
gapped mode converges to the constant dispersion $\omega(\vec{k})=2\mu$, if $U$
becomes negligibly small. As the chemical potential $\mu$ corresponds to the energy needed to add or take away one
non-interacting particle from a lattice site, this excitation may be interpreted
as the creation of a particle-hole pair.

\section{Comparison with Equilibrium Theory \label{com}}
Finally we compare our results with the ones obtained within a similar
Ginzburg-Landau theory in imaginary time 
\cite{barry}. Whereas at $T=0$ both formalisms yield identical results, a
mismatch occurs 
for finite temperature. This is surprising, since both formalisms are considered
to be equivalent in equilibrium 
\cite{chou,rammer-buch}, but there also exists a number of papers, in which
possible disagreements of the CTPF and the ITF are discussed
\cite{kobes,evans,guerin,aurenche}.

In order
to localize this mismatch, we note that the equations of motions in the ITF have
the same structure as ours in Eq.~(\ref{em2}), but the retarded and advanced
cumulants have to be replaced by the analytical continuation of the thermal
Green functions to real frequencies. For the 2-point function we find that the
analytical continuation yields exactly the retarded/advanced functions.
However, in case of the 4-point functions, the thermal Green function has a
decomposition into lower cumulants, while the retarded and advanced Green
functions do not have a similar decomposition (see Appendix A and
Ref.~\cite{grass}). Apart from this missing cumulant decomposition terms, both
formalisms agree completely, but these terms, which vanish for $T=0$, might
become large if temperature increases.

We note that the decomposition vanishes for retarded and advanced
Green functions due to the choice of our time-evolution contour in Section
\ref{keldysh}. On the Keldysh contour the path of the largest-time operator
plays no role, which results in the vanishing of contour-ordered products of 
$\Delta$-indexed operators and therefore of any possible decompositions of the
retarded and advanced Green functions. Obviously, the path of the largest-time 
operator would play a role, if not both forward and backward path were chosen to
be along the real time axis, e.g. if we shifted the backward axis by $-i\beta$
as proposed in Ref.~\cite{jakobs}. Thus we can conclude that the Keldysh
formalism working with a purely real time-evolution contour is not able to
produce the correct equilibrium configuration of the full system for finite
temperature. This result is supported by Ref. \cite{evansPRD47}, where it is
argued that throwing away the imaginary part of the time-evolution contour
reduces the generating functional $\cal Z$ to 1, whereas along the full contour
it really represents the partition function.

These formal argumentations can be backed by the physical picture that an
instantaneous change of one time-independent Hamiltonian to another
time-independent Hamiltonian should lead to the relaxation into a new
equilibrium with a new temperature. If we neglect the imaginary part of the
time-evolution contour, we formally replace the thermal average with respect 
to the new Hamiltonian by a thermal average with respect to the old,
unperturbed Hamiltonian, \textit{without} 
modifying the temperature. This might be a justified \textit{approximation} in a
non-equilibrium system where 
temperature can only be defined in the unperturbed initial state, in our
time-independent system,
however, it yields wrong results.

The agreement of both formalisms at $T=0$ can be understood as a consequence of
the Gell-Mann-Low theorem 
\cite{gell-mann} stating that the systems remains in the ground-state, if a
perturbation is adiabatically 
switched on. Since the Keldysh ansatz has pushed this switching into the
infinite past, no additional assumptions about its 
adiabatic properties had to be made \cite{rammer-buch}.
\section{Summary and Outlook \label{sum}}
We have developed a real-time and finite-temperature Ginzburg-Landau theory for
the Bose-Hubbard model by applying the 
Schwinger-Keldysh formalism of a closed time-path reviewed in Section
\ref{keldysh}. The perturbative hopping 
contributions for the free energy
have been resummed via a Legendre transformation, yielding 
a large $D$ expansion for the effective action. Therefore, we have been able not
only to 
extract the phase boundary from the equations of motion (\ref{eqm}), but also to
calculate the excitation spectra in the 
SF phase: There, the particle/hole excitations from the MI phase turn smoothly
into a gapped amplitude and a gapless phase 
mode. Surprisingly, we even obtain reliable results in the limit of small $U$,
where our theory turns into the 
Gross-Pitaevskii theory.

A comparison with a similar theory making use of imaginary times \cite{barry}, 
has shown an unexpected mismatch due to the  negligence
of the imaginary part of the time-evolution contour. In the future 
they should be overcome by a new calculation, which explicitly includes this 
imaginary path.  
At $T=0$, our theory already works in full agreement with ITF and could, thus,
be directly applied to Bose-Hubbard 
Hamiltonians with time-dependent hopping parameters $J$, which are
relevant in collapse and revival experiments \cite{Collapse1}. 
Including a possible time-dependence of $U$ might be more
problematic, but since the relevant quantity is $J/U$, we are be able 
to keep $U$ constant and put effectively
all time-dependencies into the hopping parameter \cite{schuetzhold}. Then the
only difficulty
consists in solving the equation of motion which then cannot be linearized
around an equilibrium any longer. By 
applying numerical methods, one should be able to obtain non-equilibrium results
with the formalism developed 
here.
\section*{Acknowledgement}
We acknowledge financial support from the German Academic Exchange Service
(DAAD) and from the German Research 
Foundation (DFG) within the Collaborative Research Center SFB/TR12 
\textit{Symmetry and Universality in Mesoscopic Systems}.
\appendix
\section{Calculation of the Cumulants \label{cum}}
\newcommand{\bra}[1]{\mbox{$\langle{#1}\vert$}} 	
\def\ket#1{\ensuremath{\vert{#1}\rangle}}
\def\bracket#1#2{\ensuremath{\langle{#1}\mkern1.2mu\vert\mkern1.2mu{#2}\rangle}}

As worked out in Section \ref{sem}, the equations of motion depend on the
retarded and advanced 2- and 4-point cumulants. 
Written in the Keldysh basis, the retarded and advanced Green functions are
given as averages of contour-ordered operator 
products with one operator $\op{O}^\Sigma$ and the rest being operators
$\op{O}^\Delta$. If the $\Sigma$-indexed operator 
is an annihilation operator, we have the retarded function, otherwise it is the
advanced one. From this, it can be  
directly seen
that advanced and retarded functions are linked via complex conjugation. We
thus need to calculate explicitly 
only one of both. Any cumulant decomposition of these functions necessarily
involves $\langle \op{T}_{\mathrm{c}} 
\op{O}^\Delta \op{O}^\Delta \rangle$ or $\langle \op{O}^\Delta \rangle$, both
being zero, since they contain exclusively 
$\Delta$-indexed operators. Therefore, the decomposition of the
retarded/advanced Green functions contains Green 
functions of lower order, and we don't have to distinguish between
retarded/advanced cumulants and the corresponding 
Green functions. However we should note that the unperturbed cumulants are
always local, whereas the unperturbed Green 
functions may also describe independent processes on distinct sites. In this
appendix, however, we completely suppress 
spatial variables and assume locality for all objects.

Starting with the 2-point function, we first re-write it by making use of the
Heaviside step function:
\begin{eqnarray}
 & & G^{\mathrm{R}(0)}(t_1;t_2) = i \Big\langle \op{T}_{\mathrm{c}}
\op{a}^{\Sigma}(t_1) \op{a}^{\Delta\dagger}(t_2) 
\Big\rangle_{\op{H}_0}  \\ \nonumber & &
= i\theta(t_1-t_2) \mathrm{Tr} \Big[ \mathrm{e}^{-\beta \op{H}_0}
\op{a}(t_1)\op{a}^\dagger(t_2) - \op{a}^\dagger(t_2)
\op{a}(t_1) \Big]/{\cal Z}^{(0)},
\end{eqnarray}
where ${\cal Z}^{(0)} = \mathrm{Tr} \ \mathrm{e}^{-\beta \op{H}_0}$.
The traces are best calculated in the occupation number basis solving the
eigenvalue problem
\begin{eqnarray}
 \op{H}_0 \ket{n} \equiv E_n \ket{n} \, ,
\end{eqnarray}
where the energy eigenvalues are given by
\begin{eqnarray}
\label{EV}
E_n= \frac{U}{2} n(n-1) - \mu n.
\end{eqnarray}
The retarded Green functions can thus be written as
\begin{eqnarray}
\label{GR}
&& G^{\mathrm{R}(0)}(t_1,t_2) = i \theta (t_1 - t_2) 
\sum_{n=0}^\infty \mathrm{e}^{-\beta E_n} \Big[(n+1)   
 \\ & & 
\times \mathrm{e}^{-i (E_{n+1}-E_n)(t_1-t_2)}- n \ \mathrm{e}^{-i (E_{n}-E_{n-1})(t_1-t_2)} \Big]/{\cal Z}^{(0)}
\nonumber
\end{eqnarray}
with the partition function
\begin{eqnarray}
{\cal Z}^{(0)} = \sum_{n=0}^\infty \mathrm{e}^{-\beta E_n} \, .
\end{eqnarray}
In order to determine the Fourier transform of (\ref{GR})
according to Eq.~(\ref{f-omega}), we use the integral representation of the step function:
\begin{equation}
\label{step}
\theta(t_1-t_2) = \lim_{\epsilon \rightarrow 0^+} \int_{-\infty}^{\infty} \frac{i/2\pi}{x + i\epsilon} 
\mathrm{e}^{-i (t_1-t_2)x} \mathrm{d}x.
\end{equation}
In the following we will suppress the limit-symbol for reasons of brevity.
Furthermore, we will make use of the Fourier representation of the Dirac $\delta$-function:
\begin{align}
\delta(\omega_1-\omega_2) = \frac{1}{2\pi} \int_{-\infty}^{\infty} \ \mathrm{e}^{i (\omega_1-\omega_2)t} \ \mathrm{d}t.
\end{align}
Performing the substitution $t \equiv t_1 - t_2$, we get at first
\begin{eqnarray}
&& G^{\mathrm{R}(0)} (\omega_1,\omega_2) 
= \delta(\omega_2-\omega_1) \frac{-1}{{\cal Z}^{(0)}} \sum_{n=0}^{\infty} \mathrm{e}^{-\beta E_n} 
  \nonumber \\
&& \times \int_{-\infty}^{\infty}  \frac{\mathrm{d}x}{x+i\epsilon} 
\int_{-\infty}^{\infty} \mathrm{d}t \left[(n+1) \mathrm{e}^{-i(E_{n+1}-E_n- \omega_2 +  x) t} \nonumber \right. \\
&& \left. - n \mathrm{e}^{-i(E_{n}-E_{n-1} -  \omega_2 +\hbar x) t} \right]  \, .
\end{eqnarray}
Evaluating both integrals yields then the result
\begin{eqnarray}
&& \hspace*{-8mm}G^{\mathrm{R}(0)} (\omega_1,\omega_2) 
= 2\pi  \delta(\omega_1-\omega_2) \frac{1}{{\cal Z}^{(0)}} \sum_{n=0}^\infty \mathrm{e}^{-\beta E_n}\nonumber \\
&& \hspace*{-13mm} \times \left(
\frac{n+1}{E_{n+1}-E_n - \omega_2 - i\epsilon} - \frac{n}{E_n-E_{n-1} - \omega_2 -i \epsilon} \right)\,.
\label{C2}
\end{eqnarray}
To get 
the advanced 2-point function, we must only replace the minus sign in front of the $\epsilon$ by a plus sign.

The procedure for calculating the retarded 4-point function is much the same. At first we express it in terms of Heaviside 
functions:
\begin{eqnarray}
&&\hspace*{-5mm} C^{\Sigma\Delta\Delta\Delta}(t_1,t_2;t_3,t_4) = \frac{-i}{2} \Bigg\{ 
\theta(t_1-t_2) \theta(t_2-t_3) \theta(t_3-t_4) \nonumber \\
&&\hspace*{-5mm} \times 
\Bigg\langle \Bigg[ \bigg[ \Big[ \op{a}(t_1), \op{a}(t_2) \Big], \op{a}^\dagger(t_3) \bigg], \op{a}^\dagger(t_4) \Bigg] 
\Bigg\rangle_{\op{H}_0} +\theta(t_1-t_3)
\nonumber \\
&& \hspace*{-5mm}
\times \theta(t_3-t_2) \theta(t_2-t_4)
\Bigg\langle \Bigg[ \bigg[ \Big[ \op{a}(t_1), \op{a}^\dagger(t_3) \Big], \op{a}(t_2) \bigg], \op{a}^\dagger(t_4) \Bigg] 
\Bigg\rangle_{\op{H}_0} 
\nonumber \\ 
&&\hspace*{-5mm} +
\theta(t_1-t_3) \theta(t_3-t_4) \theta(t_4-t_2)\nonumber \\
&&\hspace*{-5mm}
\times 
\Bigg\langle \Bigg[ \bigg[ \Big[ \op{a}(t_1), \op{a}^\dagger(t_3) \Big], \op{a}^\dagger(t_4) \bigg], \op{a}(t_2) \Bigg] 
\Bigg\rangle_{\op{H}_0} \Bigg\}_{t_3 \leftrightarrow t_4}
\end{eqnarray}
Here, the symbol
$t_3 \leftrightarrow t_4$ means that we still have to symmetrize the expression in these variables. 
Apart from a factor 2, it is identical to the usual definition of retarded $n$-point functions \cite{kobes}. We define
\begin{align}
\label{cr}
 2 C^{\Sigma\Delta\Delta\Delta}(t_1,t_2;t_3,t_4) \equiv C^{\mathrm{R}}(t_1,t_2;t_3,t_4).
\end{align}
Again the thermal averages can be evaluated by tracing the operator products in the occupation number basis. Then the 
resulting expression is Fourier transformed in the same way as before 
when dealing with the 2-point function. Since the number of 
different terms is much bigger now \cite{grass}, 
we do not write them down explicitly. The whole function depends only on 
time differences, thus one Fourier transformation leads to the Dirac function 
$\delta(\omega_1+\omega_2-\omega_3-\omega_4)$. From the Fourier representation 
of the three Heaviside functions, we now get three different infinitesimal
frequency shifts $\pm i \epsilon_i$. The real part of the 4-point
function is found by setting all $\epsilon_i =0$, whereas 
the imaginary part is obtained by carefully taking the limits $\epsilon_i
\rightarrow 0$. Since the knowledge of the latter is only needed for the
spectral weights but not for the spectrum itself, we only give here the real
part of the 4-point function, which has a relatively compact form:
\begin{widetext}
\begin{align}
\label{C4}
& \mathrm{Re} \ C^{\mathrm{R}}(\omega_1,\omega_2;\omega_3,\omega_4) =  - 2\pi \delta(\omega_1+\omega_2-\omega_3-\omega_4) 
\sum_{n=0}^\infty \frac{\mathrm{e}^{-\beta E_n}}{{\cal Z}^{(0)}} 
\nonumber \\& \times \Bigg\{
\frac{n(n+1)}{(E_n-E_{n-1}-\omega_1)(E_{n+1}-E_n-\omega_4)}\left(\frac{1}{E_{n}-E_{n-1}- \omega_3}-
\frac{1}{E_{n+1}-E_n- \omega_2}
 \right)
\nonumber \\& +
\frac{n(n+1)}{(E_n-E_{n-1}-\omega_3)(E_{n+1}-E_n-\omega_1)}\left(\frac{1}{E_{n}-E_{n-1}- \omega_2}-
\frac{1}{E_{n+1}-E_n- \omega_4}
 \right)
\nonumber \\& +
\frac{-(n+2)(n+1)}{(E_{n+1}-E_{n}-\omega_4)(E_{n+2}-E_{n}-\omega_1-\omega_2)}\left(
\frac{1}{E_{n+1}-E_{n}- \omega_1}+\frac{1}{E_{n+1}-E_{n}-\omega_2}
 \right)
\nonumber \\& +
\frac{n(n-1)}{(E_{n}-E_{n-1}-\omega_4)(E_{n}-E_{n-2}-\omega_1-\omega_2)}\left(
\frac{1}{E_{n}-E_{n-1}- \omega_1}+\frac{1}{E_{n}-E_{n-1}-\omega_2}
 \right)
\nonumber \\& +
\frac{2(n+1)^2}{(E_{n+1}-E_{n}-\omega_1)(E_{n+1}-E_n-\omega_2)(E_{n+1}-E_n- \omega_4)}
\nonumber \\& +
\frac{-2n^2}{(E_{n}-E_{n-1}-\omega_1)(E_{n}-E_{n-1}-\omega_2)(E_{n}-E_{n-1}- \omega_4)}
\Bigg\}_{\omega_3 \leftrightarrow \omega_4, \ \omega_2 \leftrightarrow \omega_1}.
\end{align}
\end{widetext}
If we wish to take the zero-temperature limit $\beta \rightarrow \infty$ in Eq.~(\ref{C2}) or (\ref{C4}), we must note 
that there is one occupation number $n_0$, which represents the commensurate
ground-state of the unperturbed system, and thus $E_{n_0}<E_n$ for any integer
$n\neq n_0$. By factoring out $\mathrm{e}^{-\beta E_{n_0}}$ in the Boltzmann
sums both in the denominator and the numerator, all terms in the sum with $n\neq
n_0$ remain with a factor $\mathrm{e}^{-\beta (E_n-E_{n_0})} 
\rightarrow 0$ for $\beta \rightarrow \infty$. Thus only the term with $n=n_0$
is not suppressed and survives the 
zero-temperature limit.
\end{document}